# Scaling Charitable Incentives:
# Policy Selection, Beliefs, and Evidence from a Field Experiment

**Shusaku Sasaki**[a]    **Takunori Ishihara**[b]    **Hirofumi Kurokawa**[c]

Date of Written: 31st December 2025

**Abstract**

Why are interventions with weak evidence still adopted? We study charitable incentives for physical activity in Japan using three linked methods: a randomized field experiment (N=808), a stakeholder belief survey (local government officials and private-sector employees; N=2,400), and a conjoint experiment on policy choice. Financial incentives increase daily steps by about 1,000; charitable incentives deliver a precisely estimated null. Nonetheless, stakeholders greatly overpredict charitable incentives' effects on walking, participation, and prosociality. Conjoint choices show policymakers value step gains as well as other outcomes, shaping policy choice. Adoption thus reflects multidimensional beliefs and objectives, highlighting policy selection as a scaling challenge.

[a] The University of Osaka (*Corresponding author*: ssasaki.econ@cider.osaka-u.ac.jp)
[b] Kyoto University of Advanced Science
[c] Kwansei Gakuin University

We thank Maja Adena and Steffen Huck for insightful comments and valuable discussions. We are also grateful to participants at the Science of Philanthropy Initiative 2025 Conference and the Advances with Field Experiments 2025 Conference for helpful feedback and suggestions. This research was supported by JSPS KAKENHI (Shusaku Sasaki: Grant Nos. 24K00264 and 25H00388) and by AMED (Shusaku Sasaki: Grant No. 22714343). The field experiment, the reputational evaluation survey, and the stakeholder survey experiment were all approved by the Institutional Review Board of the University of Osaka (IRB Nos. 2023CRER1215-1, 2024CRER0112, 2024CRER0127-1, and 2024CRER0127-2).
 During the preparation of this work the authors used ChatGPT to improve the readability and proofreading of the English text we have written. After using these tools, the authors reviewed and edited the content as needed and take full responsibility for the content of the manuscript.

# 1. Introduction

Even when research clearly establishes the causal effects of an intervention, such evidence does not necessarily translate into policy implementation (DellaVigna and Pope, 2018; DellaVigna et al., 2024; Hjort et al., 2021; Samek and Longfield, 2023). In practice, policymakers choose among multiple competing interventions at the policy selection stage. In this process, interventions are sometimes selected in ways that are not fully aligned with scientific evidence: interventions shown to be effective may fail to be adopted, while others with limited empirical support may be chosen. Despite the prevalence of the divergences between evidence and adoption decisions, the mechanisms underlying these divergences remain undeveloped.

Such mismatches between scientific evidence and policy adoption at the policy selection stage can, in turn, lead to the failure of policy interventions to expand and become sustained at scale. In recent years, a growing body of research—often referred to as the "science of scaling"—has systematically examined why interventions that are effective in experimental settings fail to scale when implemented in larger or different environments. The framework developed by Al-Ubaydli, List, and Suskind (2020, 2021) conceptualizes scaling as the challenge of implementing a single evidence-based intervention in a new environment, and highlights how failures of external validity can stem from limited representativeness of experimental populations and contexts, as well as from statistical and publication biases. Building on this insight, we argue that real-world policy adoption also reflects the fact that policymakers evaluate competing interventions using multidimensional objective functions that differ systematically from those typically emphasized in academic research. While researchers often focus on a narrow primary outcome, policymakers simultaneously consider other policy-relevant metrics. This perspective is consistent with and motivated by prior evidence showing that policymakers hold inaccurate beliefs about intervention effects (DellaVigna and Pope, 2018; Samek and Longfield, 2023) and that policy adoption is shaped by organizational and institutional considerations beyond experimentally measured outcomes (DellaVigna et al., 2024; Hjort et al., 2021). By empirically examining policymakers' beliefs and multidimensional objectives within a unified framework, this study brings the policy selection stage explicitly into view and contributes to the science of scaling by showing how adoption decisions may diverge from experimentally measured effectiveness even before questions of scale arise.

A canonical example of this divergence is the use of *charitable incentives*. Charitable incentives are designed to promote behavioral change by linking individual actions to charitable donations. They have been widely adopted in a variety of forms, including charity walks and runs, programs that tie blood donation to charitable giving, schemes that link event participation or consumer purchases to donations, etc. (London Marathon Events, 2024; American Cancer Society,

n.d.; Associated Press, 2023a, 2023b; ALS Association, n.d.).[1] In this study, we focus on a particular form of charitable incentives that link individuals' step counts to charitable donations, which have become widely implemented through smartphone applications and related digital platforms (Charity Miles: Walking & Running – Google Play, 2025; Apps on Google Play: Impact – Steps Fitness Charity, 2025; Big Team Challenge walking challenge statistics, 2025). Despite this widespread adoption, rigorous empirical evidence that charitable incentives generate behavioral change remains limited (Finkelstein et al., 2016; Kramer et al., 2019; Galárraga et al., 2020; Kramer et al., 2020), and furthermore there is little evidence that they consistently outperform financial incentives, for which a substantial body of evidence exists (Giles et al., 2014; Luong et al., 2021; Mitchell et al., 2013; Salmani et al., 2025; Volpp et al., 2008; Patel et al., 2016). The frequent selection of charitable incentives in practice, despite this lack of strong supporting evidence, thus provides a clear illustration of a scaling challenge that arises at the policy selection stage—namely, a divergence between scientific evidence and adoption decisions.

In the current study, we focus on this empirically salient and policy-relevant case to investigate why charitable incentives are adopted and to identify the mechanisms underlying their selection. To empirically examine this question, we conduct a series of experiments and surveys in Japan. Japan provides a particularly informative setting for studying this puzzle. Charitable incentives are widely used in practice by both private firms and local governments in Japan and appear at a frequency comparable to financial incentives in health promotion and community programs. In fact, using survey data collected for this study, 25% of local governments and 18% of private firms report having previously organized sports events featuring charitable incentives, compared with 45% and 25%, respectively, for events offering financial incentives. This institutional environment allows financial and charitable incentives to be studied as genuinely competing policy options, rather than as niche alternatives, thereby providing a natural laboratory for examining policy selection.

Why does the divergence between empirical evidence and policy selection—illustrated by the case of charitable incentives—arise in the first place? From a scaling perspective, at least two mechanisms may be at work. First, policymakers may hold inaccurate beliefs about intervention effects. Second, policymakers may optimize multidimensional objective functions that differ from those typically emphasized by researchers, placing substantial weight not only on increases in the primary behavioral outcome, but also on other outcomes such as prosociality, participation, and organizational reputation. Existing research has begun to examine these mechanisms, but has largely done so in isolation: some studies focus on policymakers' beliefs about intervention effectiveness (e.g., DellaVigna and Pope, 2018; Samek and Longfield, 2023), while others emphasize institutional and organizational factors shaping policy adoption (e.g., Hjort et al., 2021; DellaVigna et al., 2024). However, there is little work that simultaneously measures and evaluates both mechanisms within the same policy selection decision-making process.

---

[1] Large-scale charity-linked events and programs have been widely implemented across domains. The London Marathon alone raised approximately £73.5 million for charities in 2024, bringing cumulative fundraising to over £1.3 billion since its inception. The Relay For Life campaign has raised nearly $7 billion globally. Beyond event-based fundraising, consumer-linked donation schemes have also achieved substantial scale: AmazonSmile generated $449 million in donations before its discontinuation, and the viral ALS Ice Bucket Challenge raised approximately $115 million for the ALS Association in 2014 alone.

To understand why charitable incentives are frequently adopted despite limited empirical support, this study examines the divergence between scientific evidence and adoption decisions at the policy selection stage within a unified empirical framework. We combine a field experiment measuring the actual behavioral effects of financial and charitable incentives with a prediction experiment and a conjoint experiment that elicit policymakers' beliefs and multidimensional objective functions from the same set of stakeholders.

Specifically, we first establish the empirical benchmark relevant for policy selection by precisely identifying differences in behavioral effects between financial and charitable incentives (**RQ1**). We then examine whether policymakers accurately perceive these differences and how far their beliefs about intervention effectiveness diverge from experimentally measured effects (**RQ2**). Finally, we study whether charitable incentives may nonetheless be favored because policymakers place weight on outcomes beyond walking—such as prosociality, participation, and organizational reputation—and assess how strongly such multidimensional objective functions shape policy selection (**RQ3**). By addressing these three questions within a single policy selection context, this study clarifies why charitable incentives can be adopted even when their behavioral effectiveness is limited, and contributes a new perspective to the science of scaling.

To answer **RQ1**, we conducted a randomized field experiment (N = 808) in Japan in February 2024 to compare the effects of financial and charitable incentives on walking behavior. Participants were randomly assigned to a control group, a financial incentive group, a charitable incentive group, or a self-selection group in which they could choose between the two incentives, and step counts were automatically recorded via smartphone applications.

To answer **RQ2** and **RQ3**, we surveyed 2,400 stakeholders—1,200 local government officials and 1,200 private-sector employees—who are likely to be involved in planning or implementing health promotion programs. Following DellaVigna and Pope (2018) and Samek and Longfield (2023), we first elicit stakeholders' beliefs about the effects of charitable incentives on walking behavior. Importantly, we measure not only beliefs about the primary behavioral outcome, but also beliefs about additional policy-relevant outcomes, including participation, health, prosociality, and organizational reputation. We then use a conjoint experiment to estimate the relative weights that stakeholders place on these multiple dimensions when choosing among competing intervention options, thereby recovering their multidimensional policy objectives. Both components are administered to the same respondents, allowing us to study beliefs and objective functions within a unified policy selection context.

Our findings provide clear answers to the three research questions. First, the field experiment shows that financial incentives increase walking by approximately 1,000 steps on average, whereas charitable incentives generate no statistically significant effect on step counts (**RQ1**). At the same time, survey evidence indicates that charitable incentives tend to improve evaluations of the implementing organization relative to financial incentives. Second, both local government officials and private-sector employees substantially overestimate the effects of charitable incentives—not only on walking but also on multiple additional outcomes such as participation and prosociality—suggesting that inaccurate beliefs about intervention effects may distort policy selection (**RQ2**). Third, policymakers place weight not only on increases in walking but also on several additional outcomes, including participation, health, prosociality, and organizational reputation. These dimensions play an important role in policy selection, and charitable incentives are evaluated in connection with them (**RQ3**). These results indicate that the adoption of charitable incentives is less readily explained by their behavioral effectiveness and is instead more consistent with policymakers' multidimensional beliefs and values, highlighting the importance of the policy selection stage as a source of scaling challenges.

The remainder of the paper is organized as follows: **Section 2** reviews the related literature on the science of scaling and on financial and charitable incentives for behavior change. **Section 3** describes the design, implementation, and results of the randomized field experiment evaluating the effects of financial and charitable incentives on walking behavior. **Section 4** presents the stakeholder survey, including the belief-elicitation task and the conjoint experiment used to examine policymakers' beliefs and multidimensional policy objectives. **Section 5** concludes by discussing the implications of our findings for policy selection and the science of scaling.

## 2. Literature Review
### 2.1. Scaling

A growing body of research in economics examines why interventions that are effective in experimental settings often fail to scale. The frameworks in this literature—most notably Al-Ubaydli, List, and Suskind (2020, 2021)—conceptualize scaling as the challenge of implementing a single evidence-based intervention in a larger or different environment. Within this framework, failures of external validity arise from three main sources: limited representativeness of experimental populations, limited representativeness of experimental contexts, and statistical inference and publication biases. Consistent with these theoretical insights, recent empirical work documents substantial "voltage drops" when behavioral interventions are implemented in real-world administrative settings, where institutional, administrative, and operational frictions can alter the effective cost–benefit profile of the intervention (DellaVigna and Linos, 2022; Mertens et al., 2022).

A common feature of these scaling frameworks, however, is that the policy under consideration is taken as given. That is, scaling is typically framed as the decision of whether and how to expand a single intervention that has already been selected by policymakers. In practice, policymakers rarely face a binary choice between scaling and not scaling a single predetermined policy. Instead, they choose one intervention from a menu of competing options, each of which differs not only in the strength of its causal evidence but also in how well it aligns with policymakers' objectives and institutional constraints. From this perspective, failures of scaling may arise not only because treatment effects attenuate when interventions are implemented at scale, but also because policymakers may select interventions that are suboptimal with respect to the primary outcome typically emphasized in academic evaluations.

A related line of research shows that policy selection may be shaped by systematic misperceptions about policy effectiveness, even among experts and practitioners who routinely design and implement interventions. DellaVigna and Pope (2018) demonstrate that leading experts in behavioral and experimental economics frequently mispredict treatment effects across a wide range of incentivized experiments. Similarly, Samek and Longfield (2023) study fundraising practices and find that nonprofit professionals substantially overestimate the effectiveness of thank-you calls, despite large-scale field experimental evidence showing no impact on subsequent giving. These studies provide compelling evidence that inaccurate beliefs about policy effects are pervasive among decision-makers involved in policy design and selection.

Beyond beliefs about intervention effectiveness, a growing literature shows that policy adoption is shaped by organizational and institutional factors, such that rigorous experimental evidence often fails to be adopted by policymakers even before scaling becomes relevant. DellaVigna et al. (2024) document that policy adoption is strongly constrained by organizational frictions and existing administrative infrastructures, such that policymakers frequently favor interventions that fit existing practices over alternatives supported by stronger causal evidence. Hjort et al. (2021) similarly show that policymakers are responsive to experimental evidence, but that such responsiveness is shaped by institutional constraints and the compatibility of proposed policies with existing administrative structures. These studies suggest that understanding evidence-based policymaking requires distinguishing between policy selection among competing interventions and the subsequent scaling of a chosen policy as separate stages in the evidence-to-policy pipeline.

Our study builds on this literature by jointly examining policymakers' beliefs about policy effects and their multidimensional objectives within a single empirical framework, thereby explicitly incorporating policy selection into the scaling problem. We argue that real-world policy adoption reflects the fact that policymakers optimize multidimensional objective functions that differ systematically from those typically emphasized in academic research. While researchers often focus on a narrow primary outcome, policymakers simultaneously consider other policy-relevant metrics. Importantly, we measure not only stakeholders' beliefs about the effects of interventions on the primary behavioral outcome, but also their beliefs about these additional policy-relevant outcomes. We then use a conjoint experiment to estimate the relative weights that policymakers place on these multiple dimensions when choosing among competing policy options. As a result, interventions supported by stronger causal evidence may not be selected if they perform poorly along other dimensions that policymakers value. This perspective is closely related to recent calls for "Option C" thinking (Carattini et al., 2024; List, 2024), which emphasize that policy-relevant evidence must account not only for whether an intervention works, but also for whether it is likely to be adopted—a distinction we bring to the data in this paper.

**2.2. Financial and Charitable Incentives**

*Financial incentives* are a canonical policy instrument for behavior change and have been extensively studied across a wide range of domains in economics (Gneezy et al., 2011; Kamenica, 2012). Broad reviews document that monetary incentives can affect effort and behavior in settings such as labor supply, education, health behaviors, and prosocial activities, with effectiveness depending on incentive design and context. In the domain of physical activity, numerous randomized controlled trials have examined financial incentives for walking and exercise, and this evidence has been systematically synthesized in multiple reviews and meta-analyses, including comprehensive reviews and surveys in the medical and public health literatures (Giles et al., 2014; Luong et al., 2021; Mitchell et al., 2013; Salmani et al., 2025). These reviews consistently find positive short-run effects on physical activity, alongside substantial heterogeneity across studies. Importantly, this heterogeneity has itself been systematically analyzed, yielding relatively clear guidance on how effects vary with incentive design, implementation environment, and target populations. As a result, financial incentives represent the most empirically well-documented intervention for promoting physical activity, providing policymakers with a clear benchmark against which alternative incentive schemes can be evaluated (Volpp et al., 2008; Patel et al., 2016).

In contrast, empirical evidence on *charitable incentives* for increasing physical activity remains limited and fragmented. A small number of randomized trials and field studies suggest that charitable incentives can sometimes increase participation or short-run activity, but reported effects are typically modest, context-dependent, or short-lived (Finkelstein et al., 2016; Kramer et al., 2019; Galárraga et al., 2020). Other studies find no detectable effects when charitable incentives are directly compared with monetary incentives (Kramer et al., 2020). Evidence from mobile platforms further indicates that prosocial "step donation" features may increase activity among induced users, but does not establish superiority relative to financial incentives (Yuan, Nicolaides, and Eckles, working paper). Overall, this literature provides limited support for the view that charitable incentives consistently generate positive or superior effects on physical activity.

Conceptual and domain-general evidence comparing charitable and financial incentives points to similarly conditional conclusions. Imas (2014) shows that charitable incentives can elicit greater effort than financial incentives at low reward levels, but that this advantage disappears as incentive levels increase. Schwartz et al. (2021) further demonstrate that in settings where participation is voluntary, charitable incentives may reduce participation at the extensive margin by increasing moral or psychological costs. These findings suggest that the behavioral effects of charitable incentives are highly context-dependent and unlikely to dominate monetary incentives in general settings. This pattern raises a natural question: why are charitable incentives nevertheless frequently selected in practice, despite their weaker and less predictable behavioral effects?

## 3. Field Experiment
### 3.1 Design and Implementation

We conducted a field experiment in Japan to evaluate how financial and charitable incentives influence daily walking behavior. The incentives were randomly provided over a one-week period from February 19 to 25, 2024. This field experiment was approved by the Institutional Review Board of the University of Osaka (IRB nos. 2023CRER1215-1 and 2024CRER0112) and preregistered in the AEA RCT Registry on February 04, 2024, prior to the start of the intervention. A detailed pre-analysis plan was prepared as a separate document and attached to the registry entry (see Supplementary Material 1 for preregistration details).

In January 2024, we recruited smartphone users residing in the Tokyo and Osaka metropolitan areas through email invitations. Participants provided opt-in informed consent for participation and for the use of step-count data automatically recorded by their phones, mirroring the enrollment and consent procedures commonly used in real-world walking events. Before randomization, they completed a baseline survey collecting demographic and socioeconomic information, family composition, education, income, physical activity habits, subjective health, prosocial attitudes, donation experience, etc. Participants who completed the study protocol over the full experimental period received a fixed participation payment of 500 JPY, independent of the incentive scheme.

Participants were individually randomized, stratified by residential area (Tokyo / Osaka), gender (male / female), and baseline step count (high / medium / low)—into one of four groups: a control group with no incentive, a financial incentive group, a charitable incentive group, and a self-selection group. We included the self-selection group to document which incentive type participants chose when given a choice, thereby providing descriptive information on the relative attractiveness of the two incentive schemes.

Participants in the control group received a walking-promotion message but were not offered any incentives. In the financial incentive condition, participants earned 10 JPY[2] for every 1,000 steps walked, up to 12,000 steps per day (120 JPY daily, 840 JPY weekly). In the charitable incentive condition, an equivalent amount was donated to a domestic disaster-relief fund[3] based on the same step-count rule. The incentive level was deliberately set at a relatively low magnitude to ensure that financial and charitable incentives constituted a meaningful and theoretically comparable contrast (Imas, 2014). In the self-selection condition, participants chose between the financial and charitable incentive schemes after receiving identical explanations. We provided key experimental materials, including the intervention messages and the outcome measures, in Supplementary Material 2.

Participants were informed of their assignment by email on February 9, and the intervention was implemented over a one-week period from February 19 to 25, 2024. Participants received reminder messages on February 16, 19, 22, and 24 before and during the intervention week. In March 2024, participants completed a follow-up survey measuring secondary outcomes, including self-reported physical activity, subjective health, prosocial attitudes, etc.

Following the pre-analysis plan, we excluded individuals who averaged at least 12,001 steps per day in the pre-experiment, which exceeded the maximum incentivized step level, as well as those with missing step count data or incomplete survey responses. The resulting analytical sample comprised 808 participants (approximately 202 per arm), aligning with the target sample size implied by a power analysis assuming $\alpha = 0.05$, power $= 0.95$, and an expected effect size of 0.36 (Luong et al., 2021).

**3.2 Estimation and Analysis**

The analyses reported in this section follow the pre-analysis plan for the field experiment. The primary outcome is the daily number of steps automatically recorded via a smartphone app. The pre-analysis plan further specified secondary outcomes related to physical activity beyond step

---

[2] At an exchange rate of approximately 150 JPY per USD, 10 JPY is equivalent to about 6.7 cents (USD).
[3] The domestic disaster-relief fund refers to donations made through the Japanese Red Cross Society to support relief efforts following the Noto Peninsula Earthquake in Japan that occurred on January 1, 2024.

counts and subjective health, measured in the pre- and post-surveys.

Because individuals were randomly assigned to each incentive scheme, we identify average treatment effects by comparing changes in daily step counts over time across treatment and control groups. To exploit the panel structure of the data and improve statistical efficiency, we implement this randomized design using a difference-in-differences specification with individual and day fixed effects:

$$Y_{it} = \alpha + \beta_1(Financial_i \times Treated_t) + \beta_2(Charitable_i \times Treated_t) \\ + \beta_3(SelfSelect_i \times Treated_t) + \mu_i + \delta_t + \varepsilon_{it}, \quad (1)$$

where $Y_{it}$ denotes the daily step count of participant $i$ on day $t$. $Financial_i$, $Charitable_i$, and $SelfSelect_i$ are treatment indicators, and $Treated_t$ equals 1 during the intervention week and 0 otherwise. $\mu_i$ and $\delta_t$ represent individual and date fixed effects. Standard errors are clustered at the individual level to account for serial correlation.

The coefficients $\beta_1$, $\beta_2$, and $\beta_3$ capture average treatment effects for the financial, charitable, and self-selection groups, respectively. The comparison between $\beta_1$ and $\beta_2$ directly tests whether financial incentives are more effective than charitable incentives in increasing walking behavior.

Using the same difference-in-differences framework, we estimate treatment effects during the week following the intervention. We also conduct pre-specified heterogeneity analyses along a set of dimensions—weekday versus weekend, baseline step counts, gender, and region (Tokyo versus Osaka). We further examine effects on secondary outcomes.

Baseline characteristics are well balanced across the four groups. The variables examined include pre-intervention step counts, age, sex, family structure, education, income, donation experience, other physical activities, and subjective health. Detailed balance statistics are reported in **Appendix Table A.1**.

### 3.3 Results

[**Figure 1** is here.]

**Figure 1** provides a clear answer to **RQ1**: financial incentives substantially increased walking behavior, whereas charitable incentives did not. Relative to the control group, the financial incentive increases daily steps by 1,006 (p < .001), while the charitable incentive increases steps by 170 (p = 0.452). The financial incentive effect is also significantly larger than the charitable incentive effect (p < .001). The average daily step count in the control group during the baseline period was 5,743 steps, implying that the estimated financial incentive effect corresponds to approximately a 20% increase relative to baseline walking levels.

In the self-selection group, a large majority of participants (81.2%) chose the financial incentive when given a choice. This indicates that financial incentives were more attractive than charitable incentives to participants. The resulting increase in daily steps was 591 steps relative to the control group (p < .050). Although this estimate is not statistically distinguishable from the effect under mandatory financial incentives, it is notably smaller in magnitude: once participation is taken into account, the effect amounts to less than 80% of the financial-incentive benchmark. This pattern suggests that allowing choice attenuates the overall impact of incentives.

**Appendix Table A.2** shows that the main behavioral pattern is robust across prespecified subgroups. The financial incentive increases step counts in every subgroup we examine (weekday/weekend, baseline steps, gender, and region), with estimates ranging from 766 to 1,326 additional steps.[4] In contrast, the charitable incentive does not produce meaningful changes in any subgroup, with estimates consistently close to zero. The estimated effects for the self-selection group are positive on average but more heterogeneous across subgroups, with statistically significant effects concentrated among participants with lower baseline step counts.

[**Table 1** is here.]

**Table 1, Panel A** shows that the incentive schemes do not generate detectable effects on secondary outcomes beyond walking. For the pre-registered secondary outcomes, such as other physical activities and subjective health, we find no treatment effects. We also find no evidence of changes in prosocial attitudes, which were not pre-registered but are of interest given the common expectation that charitable incentives may enhance prosocial motivation: estimated effects on hypothetical donation amounts are small and statistically indistinguishable from zero across all incentive schemes.[5]

Furthermore, to study perceived legitimacy and reputational consequences of incentive schemes, we conducted a separate preregistered survey experiment with 2,400 adults in Japan.[6]

---

[4] We also find that the effect of the financial incentive is confined to the intervention period. In the week following the intervention, the estimated effects are close to zero and statistically insignificant−providing no evidence of a delayed effect of the charitable incentive (financial incentive: 23.5 steps, s.e. = 228.7, p = 0.918; charitable incentive: −204.5 steps, s.e. = 232.1, p = 0.379).

[5] Prosocial attitudes were measured in both the pre- and post-intervention surveys using a hypothetical donation task. Respondents were asked how much Japanese Yen (0–500 JPY) they would be willing to donate to support relief efforts for a large earthquake in Japan if they unexpectedly received 500 JPY.

[6] This survey experiment was preregistered independently of the field experiment and is designed to capture citizens' evaluations rather than participants' behavior or attitudes (see Supplementary Material 1 for preregistration details). We conducted this in February 2025. This experiment randomly assigned a sample of 2,400 adult respondents (mean age = 46.1 years; 50% female; broadly representative by age, gender, and region) to read a short vignette describing either a local government or a private company

They were randomly shown news about a company or a local government holding a walking event, with or without financial or charitable incentives. As reported in **Table 1, Panel B**, both incentive schemes increased citizens' favorability ratings relative to offering no incentive. For local governments (control mean = 5.38), the estimated effect was 0.51 points for the financial incentive ($p < 0.01$) and 0.90 points for the charitable incentive ($p < 0.01$), with the difference between the two effects being weakly statistically significant ($p = 0.07$). For private companies (control mean = 5.76), both incentive types increased favorability to a similar extent: 0.46 points for the financial incentive ($p < 0.01$) and 0.47 points for the charitable incentive ($p < 0.01$). These results create a wedge between behavioral effectiveness and perceived attractiveness: charitable incentives do not increase walking or prosocial attitudes in the field experiment, yet they can improve citizens' favorability especially toward local governments.

This divergence motivates our next analysis: we examine whether stakeholders' beliefs and valuations can account for program choices that depart from experimentally measured step-count impacts.

## 4. Survey and Embedded Experiments with Stakeholders
### 4.1 Design and Implementation

To complement the field experiment, this section describes a preregistered survey experiment designed to separately identify two mechanisms that may drive the adoption of charitable incentives despite limited empirical support: stakeholders' beliefs about intervention effectiveness (**RQ2**) and their multidimensional objectives in policy selection (**RQ3**). The analyses here are descriptive and interpretive in nature and are intended to shed light on stakeholders' beliefs and multidimensional policy objectives. We conducted the survey experiment between February 22 and 27, 2025. This survey experiment was approved by the Institutional Review Board of the University of Osaka (IRB no. 2024CRER0127-2) and preregistered in the Open Science Framework on February 20, 2025, prior to the start of the survey (see Supplementary Material 1 for preregistration details).

We administered the survey online through a professional research firm in Japan, which maintains a nationwide panel of Japanese respondents. The Japanese survey environment allows researchers to recruit large, well-targeted samples of organizational stakeholders while preserving demographic representativeness, and we explicitly leveraged this feature in our design. Using this infrastructure, we collected two distinct samples: 1,200 local government officials

---

organizing a walking event. Respondents were evenly split between the two organizational conditions and independently randomized to one of three incentive conditions: no incentive, a financial incentive, or a charitable incentive. Organizational favorability was measured using the average of two items capturing feelings of liking and friendliness toward the organization, each rated on a ten-point scale. The experimental design and incentive structures were constructed to mirror those used in the field experiment.

(primarily those with experience in public health, sports, cultural promotion, or urban policy departments) and 1,200 private-sector employees (mainly those with experience in human resources or corporate welfare departments).

We implemented a two-stage sampling procedure. In the first stage, we used screening questions to align age, gender, and regional composition with nationally representative distributions and to verify respondents' relevant professional experience in policy- or program-related roles. Based on these screens, we constructed two target pools—one of local government officials and another of private-sector employees. In the second stage, we administered the main survey separately to each pool, calibrating each sample to appropriate population benchmarks derived from official government statistics. This approach allows us to analyze policy selection using samples that closely approximate actual organizational decision-makers, rather than convenience samples or highly aggregated proxies.

We found that the local government officials sample comprised 1,200 respondents with substantial professional experience (mean age = 46.8 years; average tenure = 21.3 years). Managerial or senior supervisory positions accounted for 25.6% of respondents, while the majority occupied core professional roles directly involved in policy planning and implementation. A majority had experience in policy-relevant departments, particularly health and medical divisions (52.3%), alongside policy planning (30.4%), tourism or community development (29.2%), and sports or cultural promotion (24.5%). Women accounted for 67.8% of respondents, reflecting the staffing structure of health- and medical-related departments.

The private-sector employee sample also comprised 1,200 respondents with extensive experience in corporate welfare and employee benefits administration (mean age = 44.5 years; average tenure = 16.1 years). A majority were currently assigned to welfare or benefits-related departments (69.6%), while the remainder had prior experience in such roles, with an average of 7.6 years of experience in welfare-related positions. Most respondents were regular employees (73.7%), with smaller shares of part-time, contract, or temporary workers. Women accounted for 51.4% of respondents. Across both samples, respondents were drawn from a wide range of municipality sizes, from large metropolitan areas to smaller cities, indicating broad institutional and geographic coverage across local governments and firms in Japan.

Each respondent completed two main components: (1) an incentivized belief-elicitation task measuring predictions about the effectiveness of charitable incentives relative to financial incentives, and (2) a conjoint experiment measuring preferences for adopting different intervention designs. To preserve a clean separation between beliefs and preferences, the conjoint experiment was administered before the belief-elicitation task. This ordering ensured that benchmark information on the financial incentive condition did not contaminate respondents' conjoint choices.[7] We provided key experimental materials in Supplementary Material 2.

*Belief-Elicitation Task:*

Respondents participated in a financially incentivized belief-elicitation experiment following Schlag et al. (2015). The task was designed to elicit stakeholders' beliefs about the effectiveness of charitable incentives relative to financial incentives, thereby directly addressing **RQ2**. Respondents were provided with benchmark information on outcomes under financial incentives[8] from prior field and survey experiments and were asked to predict corresponding outcomes under charitable incentives for five indicators: (1) step count, (2) subjective health status, (3) prosocial attitudes, (4) number of participants willing to receive the intervention, and (5) favorability toward the implementing organization. Respondents received a monetary reward if their prediction fell within a ±5% range of the actual value.

*Conjoint Experiment:*

Each respondent also participated in a conjoint experiment designed to recover stakeholders' implicit objective functions when selecting among competing physical activity promotion programs, thereby addressing **RQ3**.[9] In each of 12 choice tasks, respondents were presented with three hypothetical intervention options and asked to select the one they would prefer their organization to adopt. Following standard practice in choice experiments (Adamowicz et al., 1994), one option was a fixed status quo (no-incentive) alternative, reflecting the realistic possibility that decision-makers may choose not to adopt any new intervention.

The remaining two alternatives in each task varied randomly across seven attributes selected to capture the key dimensions relevant to organizational decision-making in real-world walking-promotion programs: type of incentive (financial vs. charitable); expected changes in step count, health status, prosocial attitudes, participation, and public favorability (each coded as −20%, 0%, or +20%);[10] and implementation cost per participant (1,000 JPY, 3,000 JPY, or 5,000 JPY).

---

[7] The belief-elicitation survey was conducted after the completion of the field experiment, and thus analysis blinding was not implemented. This differs from DellaVigna et al. (2024), where blinding is used to address a concern specific to academic expert forecasters—namely, that the presence of a prediction task may itself signal null effects. In our setting, forecasters are practitioners rather than researchers, making such inference unlikely. Consistent with this interpretation, forecasts in our data systematically overestimate, rather than attenuate toward zero, the realized experimental effects.

[8] We provide benchmark outcomes under financial incentives for two reasons: (i) prior forecast-elicitation work (e.g., DellaVigna and Pope, 2018) provides the estimated effect of a key treatment as a benchmark for forecasters; and (ii) eliciting financial forecasts for the same five outcomes would double the prediction tasks (from five to ten). Any anchoring toward the financial benchmark appears limited: although the financial-incentive benchmark effects are essentially null for both subjective health and prosocial attitudes, Table 2 shows substantial overprediction for prosociality but not for subjective health. This pattern suggests that benchmarking is unlikely to be mechanically driving the central misbelief results.

[9] A standard conjoint design typically includes between eight and sixteen choice tasks (Bridges et al., 2011). Bansak et al. (2018, 2021) further show that even when the number of attributes or tasks increases, the estimated parameters remain stable and any deterioration in response quality is limited. In particular, they report that for designs with roughly 8–15 tasks, additional tasks do not induce substantial changes in parameter estimates (Bansak et al., 2021).

To balance cognitive burden with the ability to identify meaningful trade-offs, we followed established guidance on conjoint design (DeShazo and Fermo, 2002; Lancsar and Louviere, 2008) and implemented 12 choice tasks with seven attributes. The status quo option was fixed across tasks and represented a message-only intervention with no expected changes in any outcomes and a cost of 1,000 JPY per participant. The experimental design was generated using an orthogonal design in Ngene software and blocked into three sets of 12 tasks. Respondents were randomly assigned to one of the blocks within each sample (local government officials vs. private-sector employees).

Power calculations using the R package cjpowR (Schuessler and Freitag, 2020) indicated that, assuming power = 0.8, α = 0.05, and three attribute levels, an effective sample size of 545 respondents is sufficient to detect an AMCE of 0.03, and 1,226 respondents to detect an AMCE of 0.02. Because each respondent completed 12 tasks with two randomized alternatives, we targeted 1,200 respondents per group (2,400 in total), balancing statistical precision and budget constraints.

**4.2 Estimation and Analysis**

*Belief-Elicitation Task:*

We analyze the belief-elicitation data to evaluate accuracy. For each respondent $i$ and indicator $k$, we compare their predicted value $\hat{E}_{ik}$ for the charitable incentive to the actual experimental effect $E_k^{obs}$. Belief accuracy is defined as $|\hat{E}_{ik} - E_k^{obs}|$, and overestimation occurs when $\hat{E}_{ik} > E_k^{obs}$. We test the null hypothesis that the mean predicted effect equals the observed effect, and we also examine how much the predicted charitable incentive effect exceeds the observed financial incentive effect. Analyses are conducted separately for the two stakeholder groups (local government vs. private sector).

*Conjoint Experiment:*

To recover stakeholders' preferences over intervention attributes from the conjoint choice data, we estimate a mixed logit model grounded in a random utility framework. Attribute-specific marginal utilities are modeled as random parameters that vary across individuals, while the cost attribute is specified as a fixed parameter. To capture systematic differences in preference structures between private-sector employees and local government officials, we include interaction terms between each attribute and a private-sector dummy. Using the estimated parameters, we further compute marginal willingness-to-pay (WTP) measures for each attribute. The formal model specification and estimation procedures are presented in **Appendix B**.

---

[10] Allowing for negative as well as positive changes (−20%, 0%, +20%) across all outcome attributes reflects the possibility that interventions may have unintended or countervailing effects. For example, excessive physical activity may adversely affect health, and strong financial incentives may crowd out intrinsic or prosocial motivation by shifting attention toward reward acquisition. To accommodate such potential downsides and to avoid imposing monotonic assumptions ex ante, we designed the conjoint attributes symmetrically, permitting both improvements and deteriorations in each dimension.

### 4.3 Results

*Belief-Elicitation Task:*

We first examine whether stakeholders hold accurate beliefs about the effectiveness of charitable incentives. Respondents participated in an incentivized belief-elicitation task and received bonuses for forecasts within ±5% of the true experimental outcomes (Schlag et al., 2015).

[**Figure 2** is here.]

**Figure 2** shows our first main result: stakeholders systematically overestimate the impact of charitable incentives on walking. Local government officials predict 6,876 steps and private-sector employees predict 6,926 steps under the charitable incentive, compared with the experimentally observed mean of 5,905 steps—an overestimate of roughly 1,000 steps in both groups ($p < 0.001$ in both cases). Notably, these mean forecasts are close to (or even above) the experimentally observed mean under the financial incentive (6,740 steps), foreshadowing systematic misranking between charitable and financial incentives.

[**Figure 3** is here.]

To assess the distribution of beliefs, **Figure 3** plots the cumulative distribution functions of predicted step counts. Only 27.2% of local government officials and 23.3% of private-sector employees provide forecasts within ±5% of the true value. The distributions are shifted upward: the median prediction is 6,300 steps for officials and 6,200 steps for employees, and about half of respondents in each group predict values above the upper bound of the ±5% accuracy range (6,201 steps). Importantly, a large minority believe charitable incentives outperform financial incentives: 40.2% of officials and 36.5% of employees predict higher step counts under the charitable incentive than under the financial incentive condition (6,740 steps).

[**Table 2** is here.]

Misbeliefs extend beyond step counts. Table 2 shows that stakeholders also substantially overpredict prosociality and participation under charitable incentives, and that these misbeliefs are statistically significant. For prosociality, the experimentally observed outcome is 223 (±5% range: 221–235), whereas the mean forecasts are 254.7 among local government officials and 242.6 among private-sector employees ($p < 0.001$ for both groups)—both well outside the accuracy band and exceeding the true value by approximately 20–32 points. Misbeliefs are even larger for participation. While the actual proportion of participants selecting the charitable incentive is 18.8% (±5% range: 17.9–19.8%), local government officials predict a take-up rate of 55.2% and private-sector employees predict 53.8%, overestimating the true rate by roughly 35–36 percentage points ($p < 0.001$ for both groups).[11]

These results indicate that stakeholders not only misjudge the behavioral impact of charitable incentives on walking, but also substantially overestimate its effects on prosociality and its take-up among participants.

*Conjoint Experiment:*

[**Figure 4** is here.]

**Figure 4** reports mean willingness-to-pay (WTP) estimates for each intervention attribute among local government officials and private-sector employees, placing stakeholders' trade-offs on a common monetary scale. A basic and salient result is that the charitable incentive label itself is positively valued by stakeholders.

In the private-sector sample, the WTP for a charitable incentive (6,909 JPY) is essentially indistinguishable from that for a financial incentive (6,228 JPY). Among local government officials, the WTP for charitable incentives (4,697 JPY) is slightly lower than that for financial incentives (5,943 JPY), but the difference is not statistically significant. These results indicate that charitable incentives are broadly favored by practitioner stakeholders, independent of their demonstrated effects on walking behavior.

Beyond this baseline preference for incentive types, **Figure 4** shows that stakeholders value multiple policy-relevant objectives in addition to step-count increases, and that these valuations are quantitatively large. First, health improvements are valued more than increases in step counts in both groups: the WTP for subjective health is 3,664 JPY for local government officials and 4,503 JPY for private-sector employees, compared with 2,404 JPY and 2,646 JPY for step counts, respectively. Second, stakeholders value increasing participation: the WTP for increased participation is 1,798 JPY among officials and 2,221 JPY among employees. Third, the most salient group differences arise in prosocial and reputational dimensions. Private-sector employees value charitable incentives more (6,909 JPY) than local government officials (4,697 JPY), and they also attach a higher value to organizational favorability (1,411 vs. 757 JPY), whereas valuations of step counts are similar across groups.

---

[11] While most predicted outcomes exceed the experimentally observed values, two exceptions are worth noting. First, predictions for subjective health are slightly below the observed outcome in both groups, although the magnitude of this underestimation is small. Second, predictions for favorability differ by stakeholder type: local government officials slightly underpredict favorability toward implementing organizations, whereas private-sector employees modestly overpredict favorability relative to the observed values.

The WTP magnitudes imply that stakeholders do not choose programs to maximize step counts alone. Health improvements are valued more than step-count gains, and participation and organizational favorability are valued positively. Private-sector employees value charitable incentives and organizational favorability more highly than local government officials, while valuations of step counts are similar across groups. This preference structure implies that program choices need not track step-count impacts one-for-one; charitable incentives can be selected even when they deliver smaller increases in walking, if they bundle valued prosocial or reputational attributes.

## 5. Conclusions

Our findings provide an integrated explanation for why charitable incentives continue to be adopted in real-world health promotion programs despite limited empirical support. Our field experiment demonstrates that financial incentives clearly and robustly increased daily walking, whereas charitable incentives did not change step counts. Moreover, when participants were given a choice, take-up of the charitable incentive was limited, a pattern consistent with prior evidence that prosocial incentives often attract low participation (Schwartz et al., 2021). We also find no evidence that charitable incentives strengthened subjective health perceptions or prosocial attitudes. At the same time, charitable incentives did generate a reputational benefit: when local governments implemented charitable incentive schemes, citizens evaluated these organizations more favorably.

The stakeholder survey further reveals systematic misbeliefs about the effects of charitable incentives. Both local government officials and private-sector employees substantially overestimated their behavioral impact—not only on daily step counts, but also on secondary outcomes such as participation and prosocial attitudes. This pattern echoes and extends prior findings that practitioners often hold overly optimistic beliefs about fundraising interventions (Samek and Longfield, 2023) by showing that such misbeliefs are not limited to a single primary outcome, but also encompass multiple secondary, policy-relevant dimensions, including participation and prosociality.

Turning from beliefs to policy preferences, the conjoint experiment shows that stakeholders' program choices systematically reflect trade-offs across multiple outcomes rather than a narrow focus on step-count increases. When selecting among competing intervention options, both local government officials and private-sector employees place substantial weight on improvements in subjective health, participation, and organizational favorability, in addition to expected changes in walking behavior. As a result, interventions need not maximize walking outcomes to be selected: charitable incentives can be preferred even when their effects on step counts are limited, provided that they are perceived to perform well along other policy-relevant dimensions. This pattern is consistent with recent evidence that policy adoption depends on factors beyond experimental effectiveness (DellaVigna et al., 2024). Our contribution is to make these mechanisms explicit at the level of policy choice by directly quantifying how stakeholders trade off multiple outcomes and by revealing the beliefs and values that underlie concrete program selection decisions.

These results indicate that the adoption of charitable incentives cannot be explained by their behavioral effectiveness alone. Instead, policy selection reflects the interaction of multidimensional evaluations and stakeholders' beliefs and values regarding what constitutes policy success. Charitable incentives combine limited behavioral impact with reputational benefits, optimistic beliefs about participation and prosociality, and positive valuation of outcomes beyond step counts. In this sense, our findings suggest that multidimensional beliefs and preferences play a central role in shaping policy choices, helping to explain why interventions with weak behavioral evidence can nevertheless be adopted in practice.

At the same time, an important limitation of our study concerns the formation and direction of stakeholders' beliefs and policy objectives. Our design does not allow us to disentangle whether stakeholders emphasize outcomes beyond step counts because they hold overly optimistic or inaccurate beliefs about intervention effects, or whether they form such beliefs because they already prioritize multidimensional policy objectives. As a result, we cannot identify the causal relationship between belief formation and value-based policy preferences. Addressing this question remains an important avenue for future research.

By combining rigorous impact evaluation with insights into stakeholder perspectives, we document a disconnect between actual and perceived effectiveness. Our findings highlight the importance of a *policy-based evidence-making* approach (List, 2024), in which evidence generation is guided by policymakers' objectives, beliefs, and decision criteria rather than by behavioral effectiveness alone. Engaging with stakeholders before implementing randomized controlled trials can help clarify these objectives and ensure that research generates the types of evidence most relevant for real-world policy selection. We also emphasize the importance of communicating scientific evidence more clearly to stakeholders. Together, these efforts can foster better alignment between intervention choices and intended policy outcomes, thereby enhancing the effectiveness of public policies.

# Table and Figure

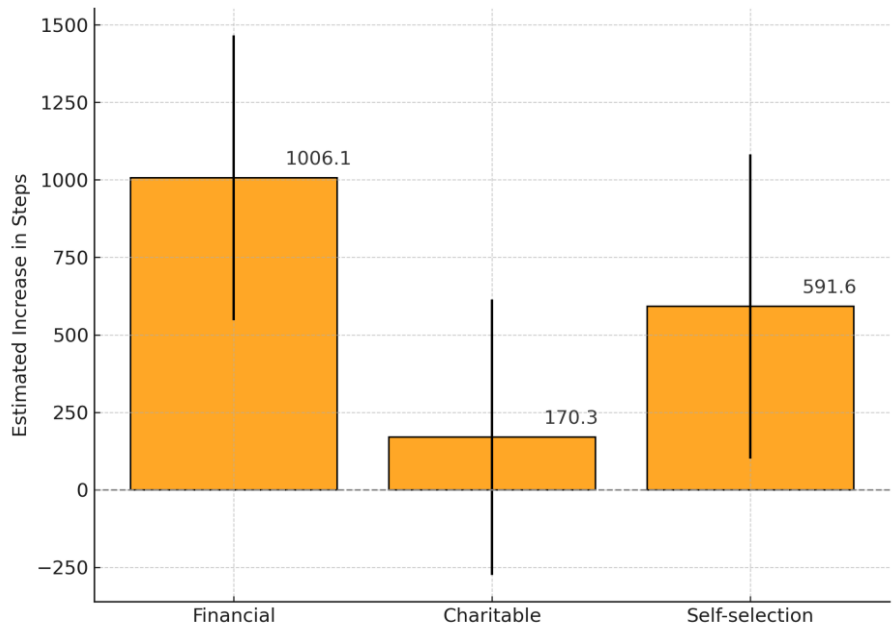

**Figure 1. Treatment Effects on Daily Step Count**

*Notes*: The figure reports estimated treatment effects on daily step counts relative to the control group from a difference-in-differences specification with individual and day fixed effects. The sample consists of 808 participants in the field experiment. Bars show point estimates for the financial incentive, charitable incentive, and self-selection conditions. Vertical lines indicate 95% confidence intervals, with standard errors clustered at the individual level. The average daily step count in the control group during the baseline period was 5,743 steps.

Financial incentives generate a sizable increase in daily steps, whereas the effect of charitable incentives is smaller and statistically indistinguishable from zero; the self-selection condition yields an intermediate effect.

**Figure 2. Actual Step Counts and Stakeholders' Beliefs under Charitable Incentives**

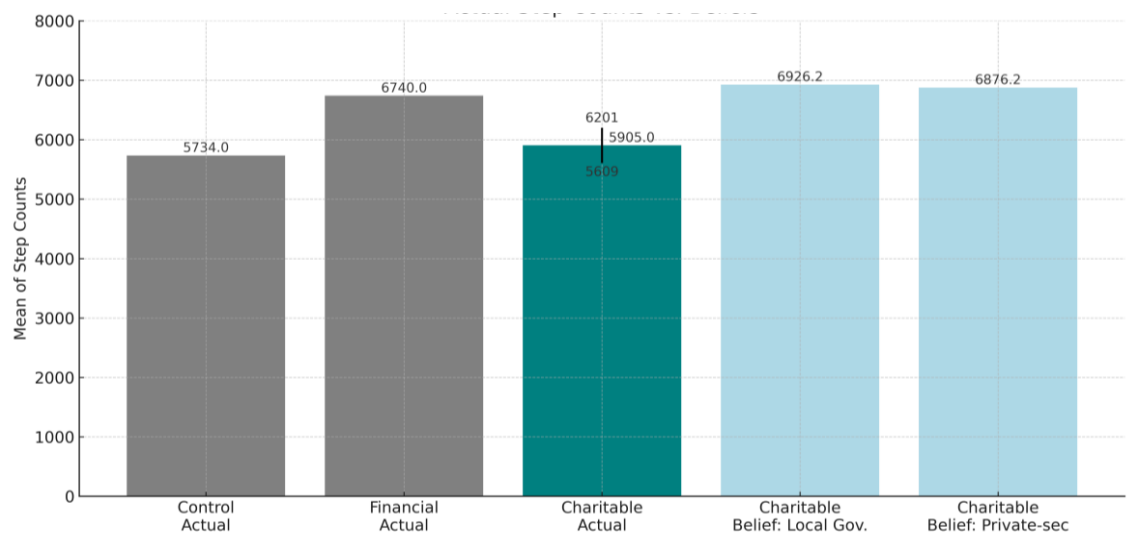

*Notes*: The figure compares experimentally observed mean daily step counts with stakeholders' predicted step counts under the charitable incentive. The dark green bar reports the observed mean for the charitable incentive condition in the field experiment, with the vertical line indicating the ±5% accuracy band used in the incentivized belief-elicitation task. Light green bars report mean predictions by local government officials (n = 1,200) and private-sector employees (n = 1,200) For reference, the figure also displays observed mean step counts under the control and financial incentive conditions.

    Stakeholders substantially overpredict step counts under the charitable incentive, with mean forecasts by both local government officials and private-sector employees exceeding the experimentally observed mean and approaching—or even surpassing—the observed mean under the financial incentive.

**Figure 3. Distribution of Stakeholders' Predicted Step Counts under Charitable Incentives**
*Panel A. Cumulative Distribution of Predicted Step Counts (Local Government Officials)*

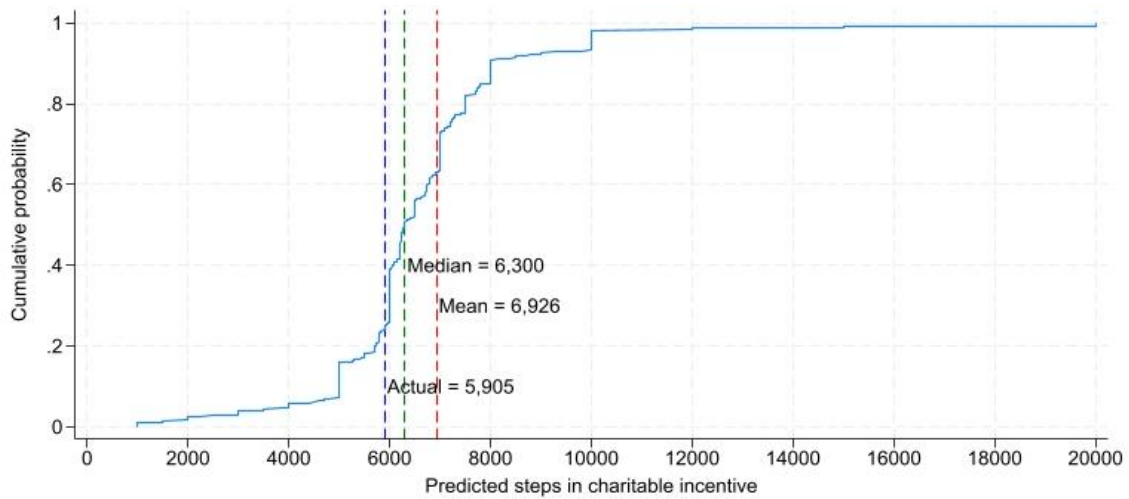

*Panel B. Cumulative Distribution of Predicted Step Counts (Private-Sector Employees)*

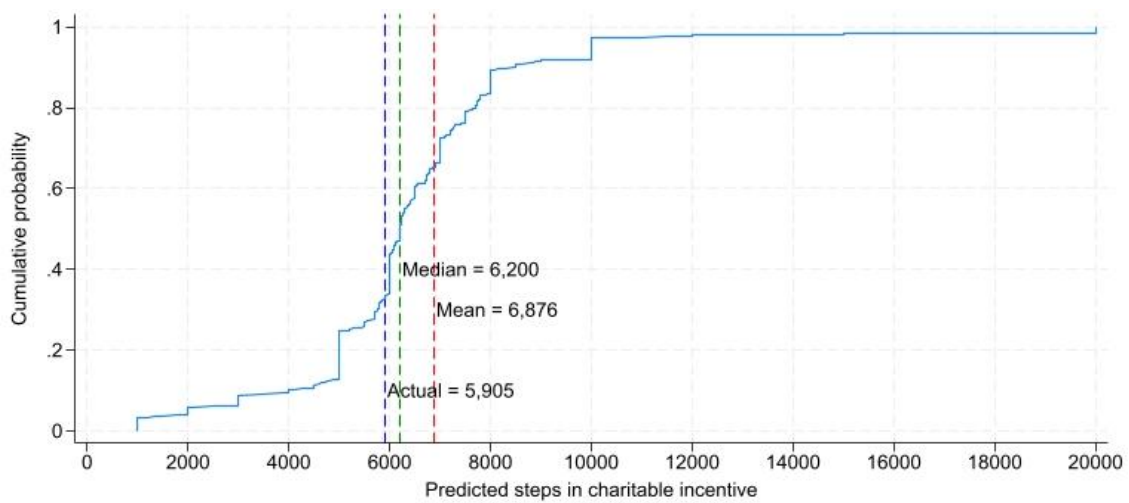

*Notes*: Panels A and B plot the cumulative distribution functions of stakeholders' predicted daily step counts under the charitable incentive. Panel A reports predictions by local government officials (n = 1,200), and Panel B reports predictions by private-sector employees (n = 1,200). The solid curve shows the empirical distribution of predicted step counts. The blue vertical line indicates the experimentally observed mean step count under the charitable incentive (5,905 steps). The green and red vertical lines indicate the median and mean, respectively, of predicted step counts in each group.

For both stakeholder groups, the distributions of predicted step counts are shifted to the right of the experimentally observed mean, indicating systematic overprediction under the charitable incentive.

**Figure 4. Willingness to Pay for Intervention Attributes**

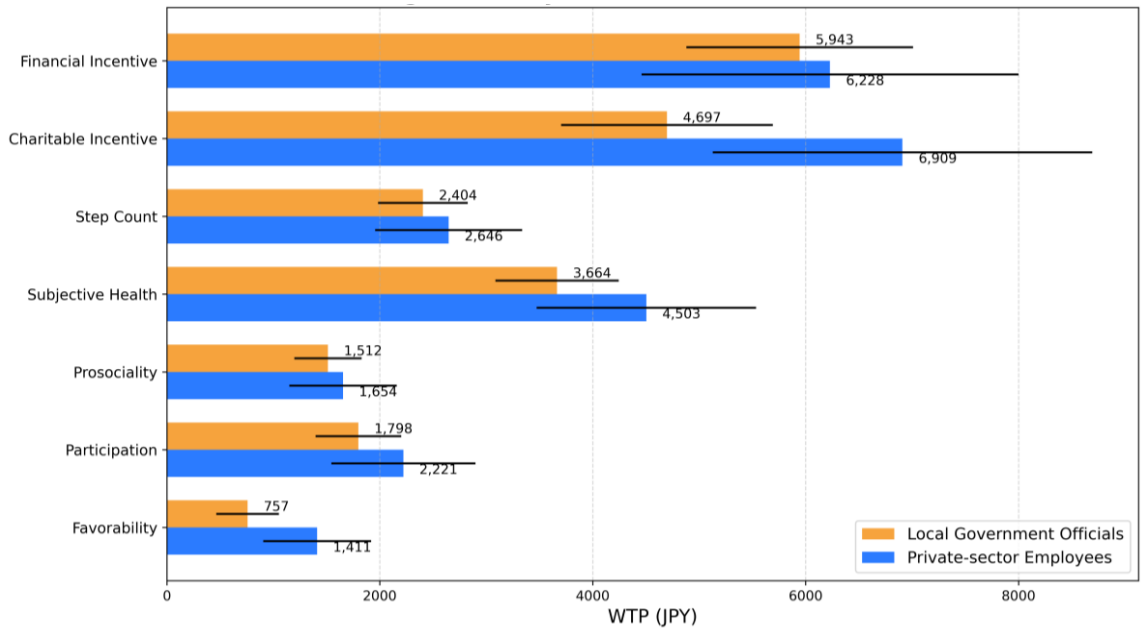

*Notes*: The figure reports willingness-to-pay (WTP) estimates for each intervention attribute derived from the conjoint experiment. WTP is expressed in Japanese yen and represents the marginal amount respondents are willing to pay for a one-unit improvement in each attribute. Estimates are shown separately for local government officials (n = 1,200) and private-sector employees (n = 1,200). Horizontal lines indicate 95% confidence intervals.

   Willingness-to-pay estimates show that stakeholders value multiple policy-relevant dimensions—including participation, subjective health, and organizational favorability—rather than focusing solely on step-count increases when evaluating intervention options.

**Table 1. Effects of Financial and Charitable Incentives
on Secondary Outcomes and Favorability**

*Panel A. Effects on Other Physical Activities, Subjective Health, and Prosociality*

| Panel A. Dependent variable: | (1) Other physical activities | (2) Subjective health status | (3) Prosociality |
|---|---|---|---|
| Intervention week | 0.00 | 0.05 | −72.18*** |
|  | (0.02) | (0.04) | (12.95) |
| Financial incentive ×Intervention week | −0.02 | −0.00 | 6.61 |
|  | (0.03) | (0.06) | (18.46) |
| Charitable incentive ×Intervention week | −0.01 | 0.02 | 5.34 |
|  | (0.02) | (0.06) | (18.20) |
| Self-selection ×Intervention week | −0.00 | 0.03 | 9.36 |
|  | (0.03) | (0.06) | (17.85) |
| Constant term | 1.25*** | 3.11*** | 287.18*** |
|  | (0.00) | (0.01) | (3.20) |
| Financial vs. Charitable (P-value) | 0.642 | 0.683 | 0.945 |
| Number of Individuals | 808 | 808 | 808 |
| Number of observations | 1,616 | 1,616 | 1,616 |

*Notes*: Panel A reports estimated treatment effects on secondary outcomes from the field experiment, estimated using a two-period difference-in-differences specification within a randomized controlled trial framework with individual fixed effects. Dependent variables are other physical activities, subjective health status, and prosociality. All specifications compare each treatment group with the control group. Standard errors are clustered at the individual level and reported in parentheses. Significance levels: *** p < 0.01, ** p < 0.05, * p < 0.10. The row "Financial vs. Charitable" reports p-values from tests of equality between the financial and charitable incentive coefficients.

Panel A shows no statistically detectable effects of financial or charitable incentives on secondary outcomes.

*Panel B. Effects on Favorability toward Implementing Organizations*

| Panel B. | (1) | (2) | (3) | (4) |
|---|---|---|---|---|
| Dependent variable: | Favorability for local governments | | Favorability for private companies | |
| Financial incentive | 0.51** | 0.50** | 0.46*** | 0.41*** |
|  | (0.20) | (0.20) | (0.10) | (0.12) |
| Charitable incentive | 0.90*** | 0.92*** | 0.47* | 0.48** |
|  | (0.18) | (0.18) | (0.24) | (0.18) |
| Constant term | 5.38*** | 4.32*** | 5.76*** | 4.06*** |
|  | (0.17) | (0.40) | (0.07) | (0.52) |
| Covariates | No | Yes | No | Yes |
| Financial vs. Charitable (P-value) | 0.070 | 0.067 | 0.968 | 0.794 |
| Number of Individuals | 1,200 | 1,200 | 1,200 | 1,200 |

*Notes*: Panel B reports estimated treatment effects on organizational favorability, estimated using ordinary least squares (OLS) regressions based on a separate reputational evaluation survey conducted in February 2025. Respondents were randomly assigned to evaluate either a local government or a private company organizing a walking event and independently randomized to one of three incentive conditions (no incentive, financial incentive, or charitable incentive). The dependent variable is organizational favorability, measured as the average of two items capturing liking and friendliness toward the organization on ten-point scales. Columns labeled "Yes" include covariates for age, gender, marital status, years of education, and household income, while columns labeled "No" exclude covariates. Standard errors are clustered at the area level and reported in parentheses. Significance levels: *** $p < 0.01$, ** $p < 0.05$, * $p < 0.10$. The row "Financial vs. Charitable" reports p-values from tests of equality between the financial and charitable incentive coefficients.

Panel B shows that both financial and charitable incentives increase organizational favorability, with larger effects for charitable incentives among local governments.

**Table 2. Misbeliefs Extend Beyond Step Counts**

|  |  | (1) | (2) | (3) | (4) | (5) | (6) |
|---|---|---|---|---|---|---|---|
|  |  | *Step counts* | *Subjective health status* | *Prosociality* | *Proportion of selecting charitable incentive* | *Favorability for local gov.* | *Favorability for company* |
| **Actual measure** | Mean | 5,905 | 3.18 | 223 | 18.81% | 6.28 | 6.23 |
|  | ±5% | [5,609 - 6,201] | [3.02 - 3.34] | [221 - 235] | [17.86 - 19.75] | [5.96 - 6.60] | [5.91 - 6.55] |
| **Predicted belief of** | Mean | 6926.35 | 3.06 | 254.66 | 55.22% | 6.11 |  |
| **local government officials** | S.D. | (5279.02) | (0.49) | (86.94) | (28.28) | (1.26) |  |
| vs. actual measure | p-value | 0.000 | 0.000 | 0.000 | 0.000 | 0.000 |  |
| **Predicted belief of** | Mean | 6876.19 | 3.00 | 242.63 | 53.79% |  | 6.36 |
| **company employees** | S.D. | (6263.02) | (0.58) | (103.93) | (30.16) |  | (1.76) |
| vs. actual measure | p-value | 0.000 | 0.000 | 0.000 | 0.000 |  | 0.011 |

*Notes*: This table compares experimentally observed outcomes with stakeholders' predicted outcomes under the charitable incentive. "Actual measure" reports the observed mean from the field experiment (or, for favorability outcomes, from the reputational evaluation survey), along with the ±5% accuracy band used in the incentivized belief-elicitation task. Predicted beliefs are reported separately for local government officials (n = 1,200) and private-sector employees (n = 1,200). Standard deviations of predicted beliefs are shown in parentheses. The rows labeled "vs. actual measure" report p-values from tests of equality between predicted beliefs and the corresponding observed outcomes.

Table 2 shows that stakeholders systematically overestimate the effects of charitable incentives not only on step counts but also on prosociality and participation, while predictions for subjective health and favorability are closer to the observed outcomes.

# Appendix A.

**Table Appendix A.1. Baseline Characteristics by Experimental Group**

|  | Sample mean by group [Standard deviation] | | | | Difference in sample means |
|---|---|---|---|---|---|
|  | *Control* | *Financial* | *Charitable* | *Self-selection* | p-value |
| Pre-step counts per day | 5743.41 | 5721.00 | 5684.39 | 5670.53 | 0.89 |
|  | [2801.39] | [2667.53] | [2666.23] | [2794.31] |  |
| Age | 49.04 | 48.53 | 48.29 | 47.59 | 0.67 |
|  | [12.23] | [11.19] | [12.32] | [11.94] |  |
| Female dummy | 0.49 | 0.49 | 0.49 | 0.49 | 1.00 |
| (1/0) | [0.50] | [0.50] | [0.50] | [0.50] |  |
| Married dummy | 0.68 | 0.75 | 0.68 | 0.72 | 0.31 |
| (1/0) | [0.47] | [0.43] | [0.47] | [0.45] |  |
| Schooling years | 15.49 | 15.27 | 15.27 | 15.36 | 0.55 |
|  | [1.81] | [1.79] | [1.77] | [1.69] |  |
| Annual income | 438.91 | 484.79 | 417.16 | 415.73 | 0.23 |
| (JPY 10,000) | [359.21] | [457.95] | [353.32] | [354.03] |  |
| No income information | 0.08 | 0.04 | 0.06 | 0.04 | 0.37 |
| (1/0) | [0.27] | [0.21] | [0.25] | [0.21] |  |
| Donation experience in 2023 | 0.42 | 0.41 | 0.37 | 0.40 | 0.80 |
| (1/0) | [0.49] | [0.49] | [0.48] | [0.49] |  |
| Frequency of physical activities | 1.23 | 1.30 | 1.24 | 1.26 | 0.33 |
| (1-5 scale) | [0.34] | [0.46] | [0.35] | [0.42] |  |
| Subjective health | 3.11 | 3.17 | 3.04 | 3.14 | 0.25 |
| (1-4 scale) | [0.70] | [0.67] | [0.65] | [0.66] |  |
| Number of individuals | 202 | 202 | 202 | 202 |  |

*Notes*: This table reports baseline characteristics of participants in the field experiment (N = 808) by experimental group. Entries show sample means, with standard deviations in brackets. The p-values in the final column are from tests of equality of means across the four groups. Pre-step counts are measured as average daily steps prior to randomization. Annual income is reported in units of 10,000JPY (150JPY ≈ 1USD). Binary variables are coded as 1 or 0, and physical activity frequency is measured on a five-point scale.

No statistically significant differences across groups are detected for any baseline characteristic.

**Table Appendix A.2. Treatment Effects on Daily Step Counts by Timing, Baseline Activity, Gender, and Area**

| | (1) | (2) | (3) | (4) | (5) | (6) | (7) | (8) |
|---|---|---|---|---|---|---|---|---|
| *Dependent variable:* | Timing: | | Baseline step counts: | | Gender: | | Area: | |
| Step counts | Weekday | Weekend | High | Low | Female | Male | Tokyo | Osaka |
| Financial incentive | 766.39*** | 1,325.79*** | 937.79*** | 1,074.49*** | 1,197.08*** | 826.21** | 944.51*** | 1,175.04*** |
| ×Intervention week | (256.62) | (368.35) | (329.36) | (322.56) | (321.69) | (337.25) | (276.48) | (432.11) |
| Charitable incentive | 81.40 | 288.95 | 60.52 | 270.42 | 234.66 | 109.74 | 318.71 | -236.29 |
| ×Intervention week | (263.21) | (340.10) | (329.44) | (298.75) | (308.10) | (329.91) | (283.90) | (331.74) |
| Self-selection | 549.85** | 647.18* | 276.68 | 920.50*** | 682.27** | 506.09 | 642.32** | 452.44 |
| ×Intervention week | (279.03) | (366.35) | (357.86) | (331.14) | (322.99) | (378.58) | (301.91) | (436.23) |
| Constant term | 5,927.68*** | 5,447.74*** | 7,789.76*** | 3,654.22*** | 5,314.72*** | 6,105.77*** | 5,693.12*** | 5,801.11*** |
| | (81.74) | (108.93) | (102.31) | (95.32) | (90.98) | (108.21) | (87.10) | (116.17) |
| Number of Individuals | 808 | 808 | 404 | 404 | 392 | 416 | 592 | 216 |
| Number of observations | 6,464 | 4,848 | 5,656 | 5,656 | 5,488 | 5,824 | 8,288 | 3,024 |

*Notes*: This table reports heterogeneous treatment effects on daily step counts across subsamples defined by timing (weekday versus weekend), baseline step counts (high versus low), gender, and residential area (Tokyo versus Osaka). The dependent variable is daily step counts. All estimates are obtained using the same difference-in-differences specification within a randomized controlled trial framework with individual and day fixed effects as in the main analysis, comparing each treatment group with the control group. Standard errors are clustered at the individual level and reported in parentheses. Significance levels: *** $p < 0.01$, ** $p < 0.05$, * $p < 0.10$.

Appendix Table A.2 shows that financial incentives robustly increase step counts across all prespecified subgroups, whereas charitable incentives generate no meaningful effects, and the effects of self selection are positive but concentrated among participants with lower baseline step counts.

## Appendix B.

We explain the estimation strategy used to recover respondents' preference structures from the online conjoint experiment. We assume the following random utility model and estimate each respondent's valuation of the intervention attributes using maximum likelihood:

$$U_{ni} = \sum_{l} \beta_n^l x_{nit}^l + \gamma m_{nit} + \varepsilon_{ni}.$$

Let $n$ index respondents, $i \in \{1,2,3\}$ denote the three alternatives presented in each choice task, and $t \in \{1,\ldots,12\}$ denote the task number. $U_{ni}$ represents the utility that respondent $n$ derives from alternative $i$ in task $t$, and $x_{nit}^\ell$ represents the level of attribute $\ell$ for alternative $i$ in task $t$. For the incentive-type attributes, we include two dummy variables: a *financial incentive* dummy and a *charitable incentive* dummy, which equal 1 when the intervention offers financial points or donation points, respectively, and 0 otherwise. For the five outcome attributes—changes in step count, health status, prosocial attitudes, number of participants, and favorability—we code the attribute levels as +1 ("increase"), 0 ("no change"), and −1 ("decrease"). The variable $m_{nit}$ denotes the implementation cost per participant. Each $\beta_n^\ell$ represents the marginal utility associated with a one-unit change in attribute $\ell$ and is modeled as a random parameter following a normal distribution. The parameter $\gamma$ captures the marginal utility of the cost attribute. In addition to the main attribute variables, the estimation includes interaction terms with a dummy variable indicating whether the respondent is employed in the private sector (1 = private-sector employee, 0 = local government official). Finally, $\varepsilon_{nit}$ is an idiosyncratic error term assumed to follow a Type I extreme value distribution.

Under these assumptions, the choice probability for respondent $n$ selecting alternative $i$ can be written as:

$$P_{ni} = \Pr(U_{ni} > U_{nj}), \qquad \forall j \neq i.$$

For the empirical analysis, we estimate a mixed logit model for the individual choice probabilities. Let

$$P_{ni} = \int \frac{\exp(V_{ni})}{\sum_j \exp(V_{nj})} f(\beta) d\beta, \qquad V_{ni} = \sum_{l} \beta_{nit}^l x_{nit}^l + \gamma m_{nit}.$$

Because the mixed logit model does not admit a closed-form likelihood, we approximate the choice probabilities using simulated maximum likelihood with Halton sequences. Using the simulated parameter draws, we compute respondents' marginal willingness to pay (WTP) for each attribute. In the random utility setting, WTP is obtained by dividing the marginal utility of each attribute by the marginal utility of money ($-\gamma$).

**Table Appendix B** reports the estimation results. First, the coefficient on the fixed parameter "cost per participant" is –0.0001 and is statistically significant at the 1% level. The interaction term with the private-sector dummy is positive and also statistically significant at the 1% level. However, because the magnitude of the interaction term is smaller than that of the main cost coefficient, the marginal utility of cost remains negative even after accounting for the interaction. This indicates that, for both local government officials and private-sector employees, interventions become less likely to be selected as their cost increases. The negative and significant cost coefficient also provides the denominator required to convert attribute utilities into the WTP measures reported below.

Next, we focus on the random parameters corresponding to the intervention types and the five outcome attributes. For local government officials, all coefficients are positive and statistically significant at the 1% level. The interaction terms between the private-sector dummy and the attribute variables are negative. Among them, the interaction terms "Private × Financial Incentive," "Private × Step Count," and "Private × Health" are statistically significant at the 1% level, while "Private × Prosociality" is significant at the 5% level. These results imply that, relative to a simple message-only intervention, interventions offering financial rewards or donation points are more preferred, and that interventions expected to increase step count, improve mental health, enhance prosocial attitudes, increase participation, or improve favorability are more likely to be selected. The negative coefficients on the interaction terms further indicate that, compared with local government officials, private-sector employees place relatively lower value on financial incentives, increases in step count, improvements in health, and increases in prosocial attitudes. At the same time, the positive main effects for these attributes remain large and statistically significant, implying that both groups still prefer interventions that deliver these improvements over those that do not.

In the mixed logit model, preference heterogeneity across individuals is reflected in the estimated standard deviations of the random parameters. Statistical significance of these standard deviations indicates the presence of meaningful variation in preferences across respondents. We observe significant heterogeneity for the attributes related to financial incentives, charitable incentives, step count, mental health, and participation. In contrast, the standard deviations for prosociality and favorability are not statistically significant, suggesting limited preference dispersion for these attributes. The interaction terms with the private-sector dummy generally do not exhibit additional heterogeneity, indicating that preference dispersion among private-sector employees is broadly comparable to that among local government officials.

Overall, these estimation results indicate that the conjoint design successfully recovers well-behaved and policy-relevant preference structures. The negative and precisely estimated cost coefficient provides a stable monetary numeraire, while the positive and statistically significant coefficients on the key program attributes confirm that respondents systematically trade off program benefits against implementation costs. Preference heterogeneity is present for several attributes, particularly in the private-sector sample. Taken together, these patterns suggest that the mixed logit model is well suited for quantifying stakeholders' multidimensional policy objectives and provide a sound empirical basis for the willingness-to-pay analyses reported in the main text.

**Appendix Table B. Random-Parameters Logit Estimates from the Conjoint Experiment**

|  | Estimate |  | s.e. |
|---|---|---|---|
| **Fixed Parameters** |  |  |  |
| Cost per participant | -0.0001 | *** | 9.76E-06 |
| Company × Cost per participant | 4.91E-05 | *** | 1.37E-05 |
| **Random Parameters** |  |  |  |
| Financial incentive | 0.8478 | *** | 0.072 |
| Charitable incentive | 0.6701 | *** | 0.068 |
| Step count | 0.3429 | *** | 0.020 |
| Subjective health status | 0.5227 | *** | 0.028 |
| Prosociality | 0.2157 | *** | 0.018 |
| Participation | 0.2565 | *** | 0.021 |
| Favorability | 0.1080 | *** | 0.020 |
| Company × Financial incentive | -0.2651 | *** | 0.103 |
| Company × Charitable incentive | -0.0236 |  | 0.098 |
| Company × Step count | -0.0954 | *** | 0.028 |
| Company × Subjective health status | -0.1014 | *** | 0.039 |
| Company × Prosociality | -0.0610 | ** | 0.025 |
| Company × Participation | -0.0487 |  | 0.030 |
| Company × Favorability | 0.0240 |  | 0.029 |
| **Standard Deviations of Random Parameters** |  |  |  |
| Financial incentive (sd) | 2.2529 | *** | 0.050 |
| Charitable incentive (sd) | 2.1470 | *** | 0.049 |
| Step count (sd) | 0.1151 | ** | 0.055 |
| Subjective health status (sd) | 0.6580 | *** | 0.023 |
| Prosociality (sd) | 0.0527 |  | 0.038 |
| Participation (sd) | 0.2818 | *** | 0.038 |
| Favorability (sd) | 0.0001 |  | 0.032 |
| Company × Financial incentive (sd) | 0.0108 |  | 0.157 |
| Company × Charitable incentive (sd) | 0.3166 | * | 0.171 |
| Company × Step count (sd) | 0.0767 |  | 0.071 |
| Company × Subjective health status (sd) | 0.0096 |  | 0.067 |
| Company × Prosociality (sd) | 0.0337 |  | 0.051 |
| Company × Participation (sd) | 0.1270 |  | 0.157 |
| Company × Favorability (sd) | 0.0355 |  | 0.045 |
| N | 2,400 |  |  |
| obs | 28,800 |  |  |
| Pseudo R^2 | 0.1620 |  |  |
| LRI | -26514.94 |  |  |

*Notes*: This table reports parameter estimates from a random-parameters logit model estimated using data from the conjoint experiment. The sample consists of 2,400 respondents, each completing 12 choice tasks, yielding 28,800 observations. Fixed parameters capture average effects common across respondents, while random parameters allow for unobserved heterogeneity in preferences across individuals. Interaction terms with "Private" indicate differences in preferences between private-sector employees and local government officials. Standard errors are reported in the final column. Significance levels: *** $p < 0.01$, ** $p < 0.05$, * $p < 0.10$.

# Supplementary Materials I
# Pre-registration and Pre-analysis Plan

## Pre-registration Overview
## A.1 Field Experiment

The field experiment reported in this paper was preregistered in the AEA RCT Registry prior to the start of the intervention.

- Registry: AEA RCT Registry (AEARCTR-0012944)
- Registration date: February 04, 2024
- Trial title: *Financial and Prosocial Incentives for Physical Activity: A Field Experiment*
- Intervention period: February 19–25, 2024
- Primary outcome:
  Daily step counts automatically recorded via participants' smartphone applications
- Secondary outcomes (prespecified):
  Other physical activities; subjective health status (self-reported)
- Additional outcomes (not prespecified):
  Prosocial attitudes (reported as exploratory)
- Randomization:
  Individual-level stratified randomization by region (Tokyo/Osaka), gender, and baseline step counts
- Prespecified treatment arms:
  Control; financial incentive; prosocial incentive; self-selection
- Estimation strategy:
  Difference-in-differences with individual and day fixed effects and participant-level clustered standard errors
- Deviations from the pre-analysis plan:
  None with respect to the primary outcome, prespecified secondary outcomes, sample construction, or estimation strategy

## A.2 Reputation Survey Experiment

The reputation survey experiment reported in Section 3.3 was preregistered independently of the field experiment and is designed to examine citizens' evaluations of organizations implementing incentive-based walking programs.

- Registry: AEA RCT Registry (AEARCTR-0015374)
- Registration date: February 12, 2025
- Trial title: *A Survey Experiment on Citizens' Perceptions and Impressions of Corporate and*

*Local Government Health Promotion Initiatives*

- Sample:
  2,400 adult respondents in Japan (1,200 assigned to company news; 1,200 assigned to local government news)
- Design:
  Between-subject survey experiment with randomized exposure to hypothetical news articles describing walking events with no incentive, a monetary incentive, or a charitable incentive
- Primary outcome:
  Favorability toward the implementing organization (average of two liking-related items measured on a 10-point scale)
- Secondary outcomes:
  None prespecified
- Randomization:
  Individual-level stratified randomization by age, sex, and area of residence
- Prespecified hypotheses:
  Monetary and charitable incentives increase favorability relative to no incentive, for both companies and local governments
- Estimation strategy:
  Ordinary least squares comparisons across experimental conditions with prespecified covariates and clustered standard errors
- Role in the paper:
  Supplementary evidence on perceived legitimacy and reputational consequences of incentive schemes, intended to complement the behavioral results from the field experiment
- Deviations from the pre-analysis plan:
  None with respect to the experimental design, primary outcomes, or main estimation strategy.

## A.3 Stakeholder Survey Experiment

The stakeholder survey experiment reported in Section 4 was preregistered on OSF and is designed to elicit stakeholders' beliefs and preferences regarding incentive-based health promotion programs. This survey is intended to complement the field experiment by examining how policy-relevant beliefs and valuations relate to experimentally measured behavioral effects, rather than to estimate causal treatment effects.

- Registry: OSF (Open Science Framework)
- Registration title: *Conjoint Analysis Of Stakeholder Preferences For Company And Local*
- *Government Health Promotion Initiatives*
- Registration date: February 20, 2025

- Planned study period:

  Survey start date: February 22, 2025

- Sample:

  Local government officials (1,200 respondents) and private company employees (1,200 respondents), recruited through a nationwide online survey panel. Respondents were screened to match nationally representative distributions by age, sex, and residential region within each stakeholder group.

- Design:

  Survey experiment consisting of two main components:

  (i) an incentivized belief elicitation task measuring respondents' predictions about the effects of charitable incentives on multiple outcome dimensions.

  (ii) a conjoint experiment eliciting preferences over alternative incentive-based programs, and

- Conjoint attributes:

  Incentive type (financial or charitable), expected changes in step count, health status, prosocial attitudes, number of participants, favorability toward the implementing organization, and implementation cost per participant. Each respondent completed 12 choice tasks, choosing among two randomly generated programs and a fixed outside option without incentives.

- Primary outcomes:

  Program choice in the conjoint experiment and stated beliefs about the effects of charitable incentives on step count, health status, prosocial attitudes, participation, and organizational favorability.

- Randomization:

  Individual-level random assignment to conjoint task blocks generated using an orthogonal design.

- Estimation strategy:

  Average marginal component effects (AMCEs) for conjoint attributes and descriptive analyses of belief accuracy and correlations between beliefs and stated valuations. Analyses follow the prespecified survey design and estimation approach.

- Role in the paper:

  Evidence on how stakeholders' beliefs and valuations may rationalize the adoption of incentive schemes that differ from those identified as behaviorally effective in the field experiment.

- Deviations from the pre-analysis plan:

  None with respect to the survey design, primary outcomes, or main analytical approach.

# Supplementary Materials I
# Experimental Materials and Survey Instruments

## Supplementary Materials B. Field Experiment
### B.1 Experiment Schedule

| Period / Date | Description |
| --- | --- |
| **December 20–25, 2023** | Recruitment of participants and collection of informed consent. |
| **January 22–28, 2024** | Administration of the baseline survey. |
| **February 9, 2024** | Group-specific notifications were sent. Participants in the control group received a simple message encouraging walking, while those in the self-selection group were asked to choose online between a financial incentive and a charitable incentive. |
| **February 16, 19, 22, and 24, 2024** | Reminder notifications were sent to participants in the intervention groups only. |
| **February 19–25, 2024** | Intervention period: the walking event was implemented. |
| **March 11–15, 2024** | Administration of the post-intervention survey. |

**B.2 Incentive Messages Shown to Participants**

**B.2.1 Financial Incentive Message**

This image presents the financial incentive condition shown to participants during the field experiment. The message informs participants that, during the one-week intervention period (February 19–25), they will receive a monetary reward of 10 Japanese yen (JPY) for every 1,000 steps walked. The incentive is framed as points equivalent to cash rewards.

The message also provides a concrete example to clarify the incentive structure: if a participant walks 10,000 steps per day, they can earn an additional 70 JPY per day, corresponding to the incremental rewards accumulated in excess of the baseline. The visual design emphasizes monetary gains using images of coins and explicit numeric amounts, highlighting the direct financial benefit of increased walking.

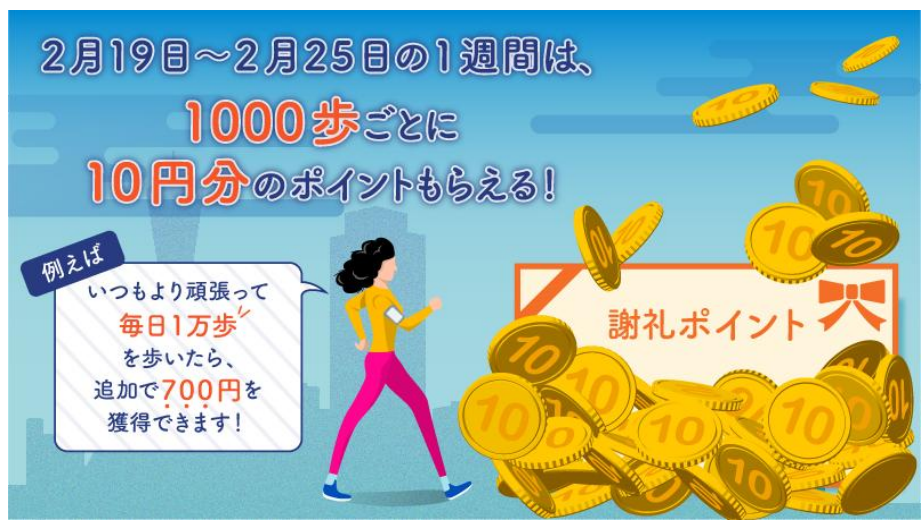

**B.2.2 Charitable Incentive Message**

This image presents the charitable incentive condition shown to participants during the field experiment. The message informs participants that, during the one-week intervention period (February 19–25), a donation of 10 Japanese yen (JPY) will be made for every 1,000 steps they walk. Unlike the financial incentive condition, the monetary reward does not accrue to the participant personally but is donated to a designated charitable cause.

To clarify the incentive structure, the message provides a concrete example: if a participant walks 10,000 steps per day, a total of 70 JPY per day will be donated on their behalf. The image visually emphasizes the donation mechanism by depicting coins being placed into a transparent donation box labeled as a disaster relief fund for *the 2024 Noto Peninsula earthquake*. The framing highlights the prosocial nature of the incentive by making the charitable recipient salient while keeping the monetary amount identical to that in the financial incentive condition.

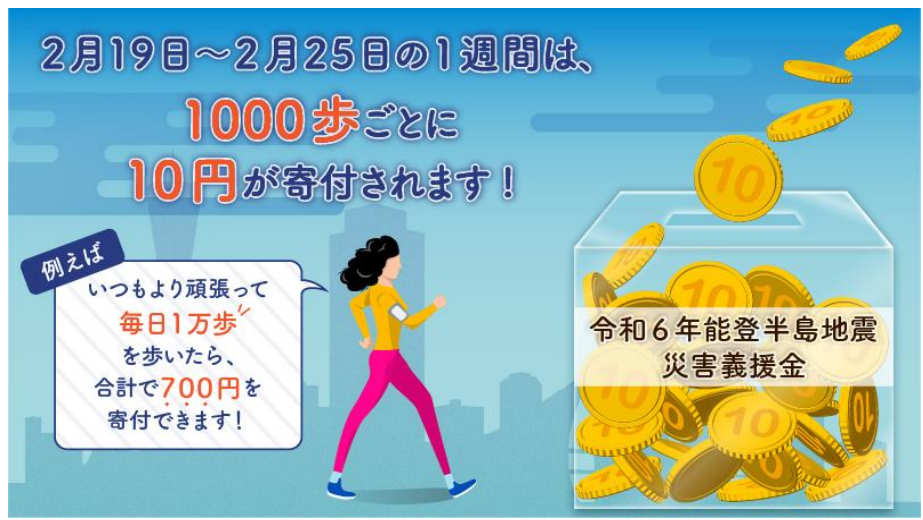

**B.3 Secondary Outcome Measures**

The following questions were used to measure secondary outcomes. These outcomes were measured using identical survey items administered both before and after the intervention period.

**B.3.1 Other Physical Activities**

Respondents were asked about the frequency with which they usually engage in various types of physical activities other than walking.

**Question:**

*How often do you usually engage in the following physical activities?*

**Activity categories (asked separately for each):**
- Cycling
- Sports gym or swimming
- Aerobics or yoga
- Ball sports
- Other physical activities

**Response options (for each activity):**
1. Four or more times per week
2. Two to three times per week
3. Once per week
4. One to three times per month
5. Do not engage in this activity

*Responses were reverse-coded, summed across the five activity categories, and divided by five to construct an average index, with higher values indicating more frequent physical activity.

**B.3.2 Subjective Health Status**

Respondents were asked to report their current subjective health status.

**Question:**

*How would you rate your current health status?*

**Response options:**
1. Good
2. Fairly good
3. Not very good
4. Poor

*Responses were reverse-coded so that higher values indicate better self-reported health and used directly as the outcome variable.

### B.3.3 Prosociality: Hypothetical Donation Measure

To measure prosocial behavior, respondents were asked a hypothetical donation question.

**Question:**

*Suppose that you win a free lottery and receive 500 points (1 point = 1 Japanese yen). Out of these 500 points, how many points would you be willing to donate to support relief efforts and assistance for victims of the 2024 Noto Peninsula earthquake?*

**Response format:**

Respondents entered an integer between 0 and 500.

*The reported donation amount (0–500 points) was used directly as the outcome variable without transformation.

## Supplementary Materials C. Reputation Survey Experiment

**C.1 Common Instruction**

Before reading the news article, all respondents were shown the following instruction:

*"Please read the following news article. This is a hypothetical article. Please imagine that the situation described is real while reading it. You may proceed to the next page after at least five seconds."*

After this instruction screen, respondents were randomly assigned to one of the news articles described below.

**C.2.1 Control Condition (Local Government Version)**

In February 2024, a local government in Japan organized a one-week walking event to promote citizens' health. During this one-week period, participating citizens received messages encouraging them to walk more, such as:

*"Walking is an exercise that you can start today. Why not try walking a little more than usual?"*

**\*Differences in the Private Company Version**

The private company version was identical to the local government version, except that references to the local government and citizens were replaced with references to the private company and employees. No other differences existed.

**C.2.2 Financial Incentive Condition (Local Government Version)**

In February 2024, a local government in Japan organized a one-week walking event to promote citizens' health. During this one-week period, participating citizens received messages encouraging them to walk more, such as:

*"Walking is an exercise that you can start today. Why not try walking a little more than usual?"*

In addition, during the event period, for every 1,000 steps walked, participating citizens received 10 Japanese yen worth of points that could be used for online shopping. These points were provided by organizations sponsoring the event.

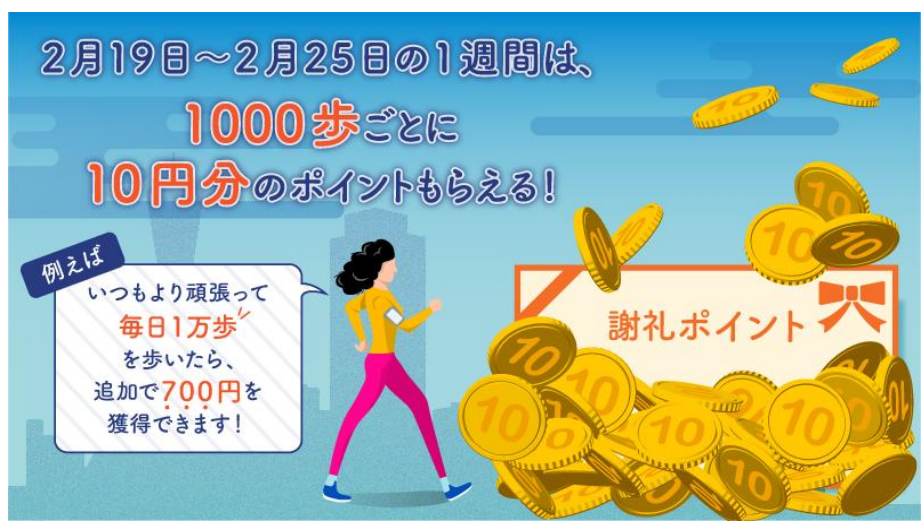

**\*Differences in the Private Company Version**

The private company version was identical to the local government version, except that references to the local government and citizens were replaced with references to the private company and employees, and the financial incentives were described as being provided directly by the company rather than by event sponsors. No other differences existed.

**C.2.3 Charitable Incentive Condition (Local Government Version)**

In February 2024, a local government in Japan organized a one-week walking event to promote citizens' health. During this one-week period, participating citizens received messages encouraging them to walk more, such as:

*"Walking is an exercise that you can start today. Why not try walking a little more than usual?"*

In addition, during the event period, for every 1,000 steps walked, 10 Japanese yen were donated on behalf of each participating citizen to disaster relief activities supporting victims of the 2024 Noto Peninsula earthquake. The donations were made by organizations sponsoring the event.

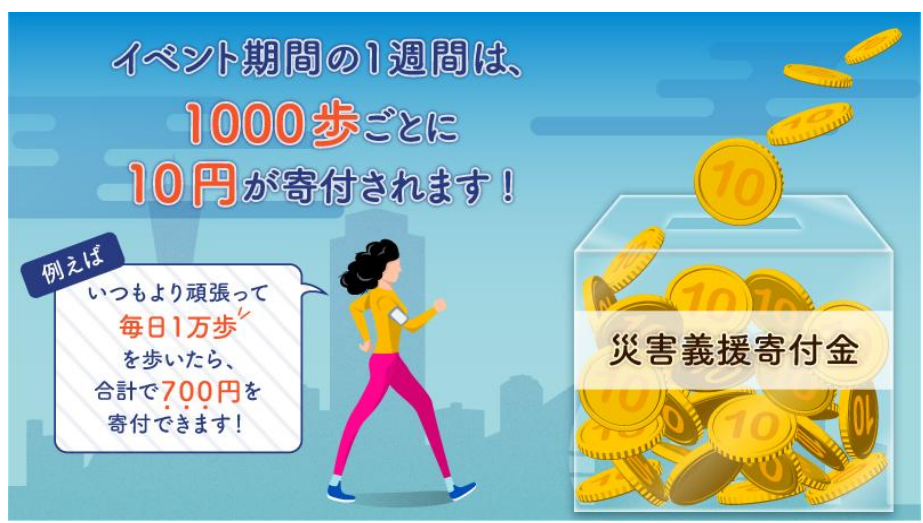

**\*Differences in the Private Company Version**

The private company version was identical to the local government version, except that references to the local government and citizens were replaced with references to the private company and employees, and the donations were described as being made directly by the company rather than by event sponsors. No other differences existed.

**C.3 Outcome Measures: Favorability Toward the Implementing Organization**

After reading the news article, respondents were asked to evaluate the implementing organization (a local government or a private company) on the following affective dimensions:

"After reading the news article, how strongly did you feel the following emotions toward the local government (or the private company)?"

**Items**
- **Liking**
- **Friendliness**

**Response Scale**

Both items were measured on a 10-point Likert scale, where:
- **1 = Not at all**
- **10 = Very strongly**

Responses to the two items were averaged to construct a **favorability index**, which served as the primary outcome variable in the reputation survey experiment. Higher values indicate more favorable evaluations of the implementing organization.

# Supplementary Materials D.
# Stakeholder Survey and Embedded Experiments

**D.1 Conjoint Experiment: Policy Preferences over Incentive-Based Programs**

**D.1.1 Introduction and Scenario**

Before starting the choice tasks, respondents were shown the following instruction and scenario.

"Please imagine the situation described below and answer what you would do if you were in this position. You may proceed to the next page after at least 20 seconds."

Respondents were then asked to consider the following hypothetical situation:

*In your local government, you are considering organizing a one-week walking event to promote citizens' health. You are a staff member involved in planning and designing this event. Specifically, the following two policy options are being considered:*

**Policy Option A: "Message + Financial Points"**

- During the one-week event period, participating citizens would regularly receive messages on their smartphones stating:

    *"Walking is an exercise that you can start today. Why not try walking a little more than usual?"*

- In addition, during the event period, participating citizens would receive **10 Japanese yen worth of points** for every **1,000 steps** they walk. These points could be used for online shopping.

- *The points would be funded by voluntary organizations sponsoring the event.*

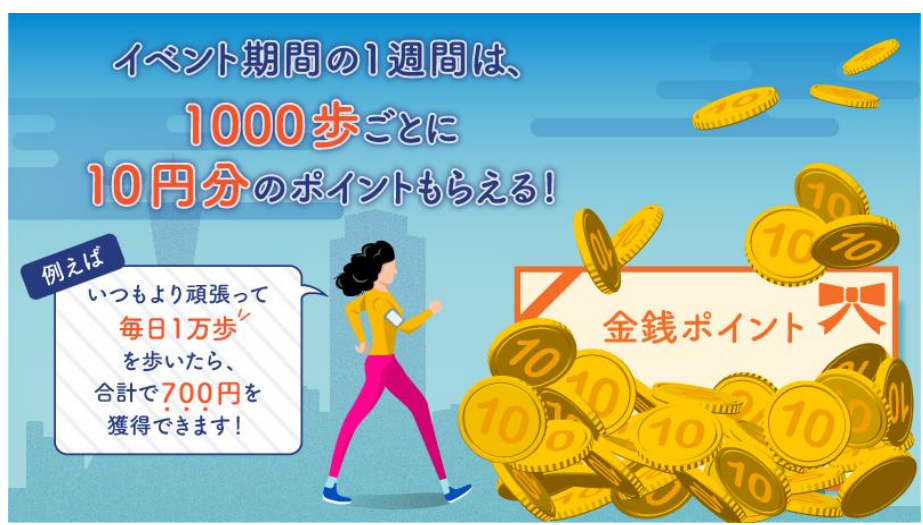

**Policy Option B: "Message + Donation Points"**

- During the one-week event period, participating citizens would regularly receive the same messages on their smartphones:

    *"Walking is an exercise that you can start today. Why not try walking a little more than usual?"*

- In addition, during the event period, **10 Japanese yen** would be donated **on behalf of each participating citizen** for every **1,000 steps** they walk, to support disaster relief activities related to natural disasters occurring in Japan.

- *The donations would be funded by voluntary organizations sponsoring the event.*

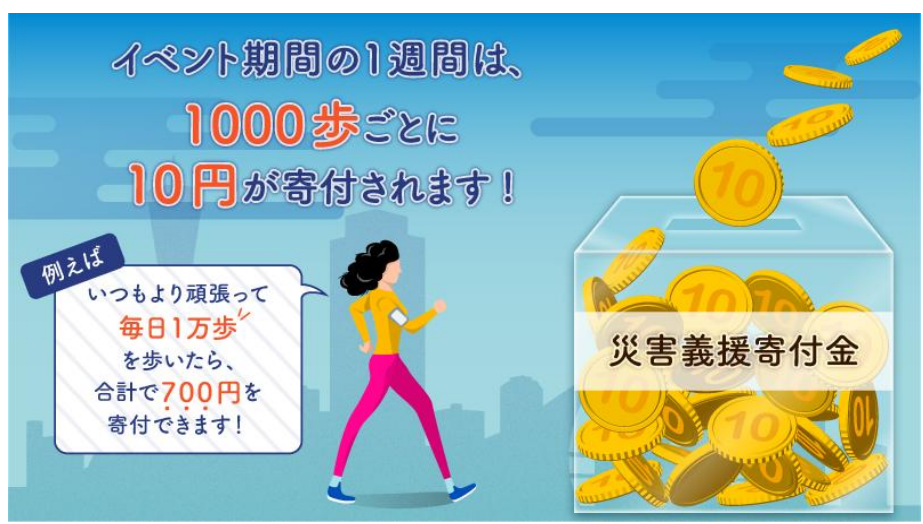

From the next page onward, you will be shown two different policy options at a time, together with information on the following six attributes. As the person responsible for planning this event, please choose which policy you consider preferable.

**Attributes Shown in Each Choice Task**

1. Expected change in **participants' step counts**
2. Expected change in **participants' health status**
3. Expected change in **participants' prosocial attitudes**
4. Expected change in the **number of event participants**
5. Expected change in **public liking toward your local government** when people across Japan see news about the policy
6. **Cost per participant** that your local government would need to bear to implement the policy

    *(including event operation costs and costs related to measuring outcomes such as step*

*counts, health status, and prosocial attitudes)*

**\*Differences in the Private Company Version (Conjoint Experiment)**

The private company version of the conjoint experiment was identical to the local government version, except that references to the local government and citizens were replaced with references to the private company and employees, and respondents were instructed to answer as planners designing a walking event for their own company. No other differences existed.

## D1.2. Example of a Conjoint Choice Card

Please read the following information carefully and choose **which program you prefer: Program A or Program B**.

|  | Program A | Program B | Neither (Support Program C) |
|---|---|---|---|
| **Program description** | **"Message + Donation Points"** For every 1,000 steps walked by a citizen, 10 JPY is donated on behalf of the citizen to disaster-relief activities for victims of natural disasters in Japan. | **"Message + Financial Points"** For every 1,000 steps walked by a citizen, the citizen receives 10 JPY worth of points that can be used for online shopping. | **"Message only"** Periodic messages encouraging walking are sent. (No points are provided.) |
| **Effect on participants' daily step count** | No change | Increase (+20%) | No change |
| **Effect on participants' health status** | Improve (+20%) | Worsen (−20%) | No change |
| **Effect on participants' motivation to donate for disaster relief** | Increase (+20%) | Decrease (−20%) | No change |
| **Effect on the number of event participants** | Increase (+20%) | Increase (+20%) | No change |
| **Effect on public favorability toward your local government (when people nationwide see news about the policy)** | Increase (+20%) | Increase (+20%) | No change |
| **Cost per participant borne by your local government** | 3,000 JPY | 1,000 JPY | 1,000 JPY |

Which program do you prefer, **Program A or Program B?**

1 **Support Program A**
2 **Support Program B**
3 Support neither program (support Program C)

**D.2 Belief-Elicitation Task: Stakeholders' Expectations about Policy Effects**

**D.2.1 Instruction Screen**

"From this point onward, you will be asked to *predict* how policies that provide financial points or donation points affect participants' outcomes—such as step counts, health status, and prosocial attitudes—in an actual field experiment.

Depending on the accuracy of your predictions, you will have a chance to earn additional reward points worth up to several thousand yen. Please read the following information carefully and answer the questions."

*(You may proceed to the next page after at least 20 seconds.)*

**D.2.2 Overview of the Field Experiment (Benchmark Study)**

Respondents were then provided with the following overview of the field experiment conducted by a research team at the University of Osaka in **February 2024**:

- Participants were members of the general public aged **20 or older**, residing in the **Greater Tokyo Area** (Tokyo, Kanagawa, Chiba, Saitama) or the **Kansai Area** (Osaka, Kyoto, Hyogo, Nara).
- Participants were randomly assigned to **three groups** of **202 participants each** (Groups 1–3).

**Group 1 (Message only).** During the one-week event period (**February 19–25, 2024**), participants received messages encouraging them to walk more than usual.

**Group 2 (Message + Financial points).** In addition to receiving the messages, during the event period (**February 19–25, 2024**), participants received **10 Japanese Yen worth of points** (usable for online shopping, etc.) for every **1,000 steps** walked. The points were provided by the research project team.

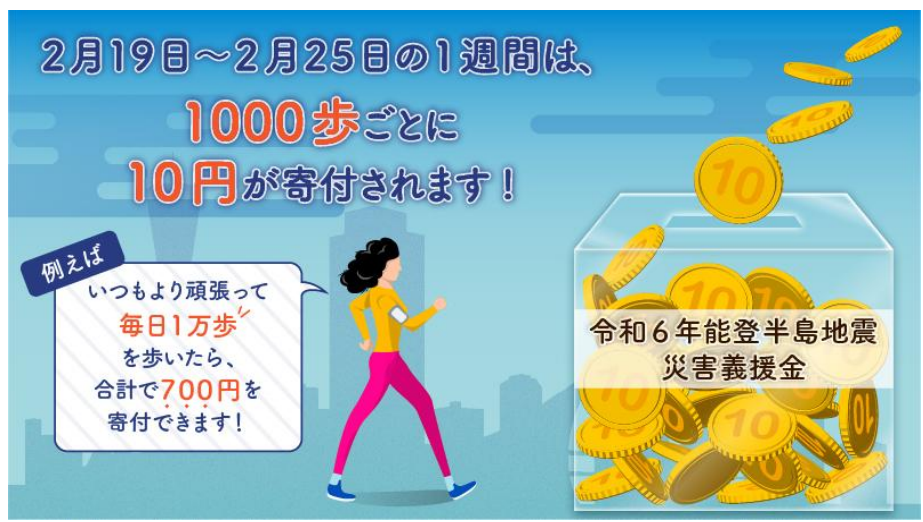

**Group 3 (Message + Donation points).** In addition to receiving the messages, during the event period (**February 19–25, 2024**), **10 Japanese yen** were donated **on behalf of each participant** for every **1,000 steps** walked, to support disaster relief activities for victims of the **2024 Noto Peninsula earthquake**. The donations were made by the research project team.

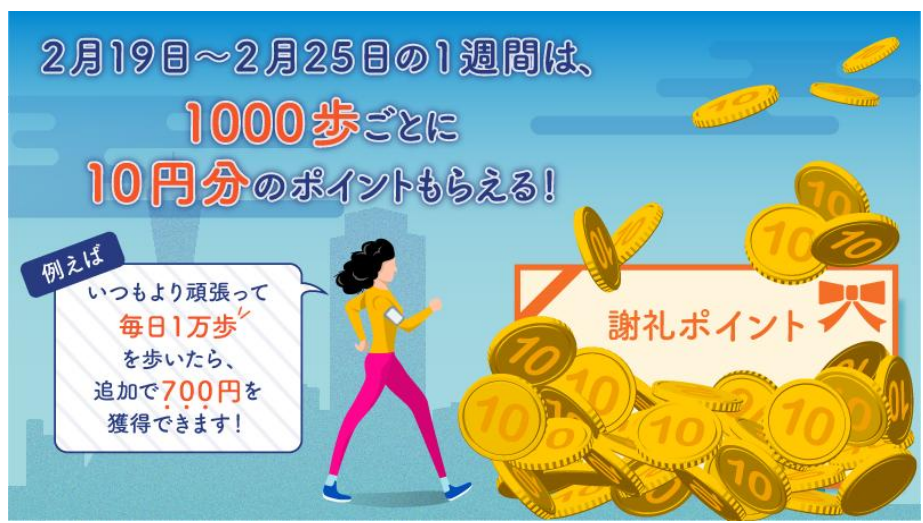

Respondents were then asked to predict, as accurately as possible, the **average daily step count during the event period** for **Group 3 (Message + Donation points)**.

**D.2.3 Results Shown to Respondents (Benchmark Information)**
(You may proceed to the next page after at least 15 seconds.)
Respondents were shown summary results for the pre-event period and for Groups 1 and 2 during the event period:

**Pre-event period (January 15–21):**
- There were no differences in step counts across Groups 1–3 in the month prior to the event.
- Across all groups, participants walked approximately **5,700 steps per day** on average.
- The groups were similar in composition (e.g., gender balance, average age, and residential area).

**Event period (February 19–25):**
- **Group 1 (Message only):** Average daily steps during the event period were **5,734 steps**
  - Compared with the pre-event month, this was nearly unchanged (a **34-step increase** in levels).
- **Group 2 (Message + Financial points):** Average daily steps during the event period were

**6,740 steps**
- This was **1,006 steps higher** than Group 1 (5,734 steps).

Respondents were then asked:

**"Please predict, as accurately as possible, the average daily step count during the event period for Group 3 (Message + Donation points)."**

**D.2.4 Incentive Rule for the Belief-Elicitation Task**

Respondents were informed that they would have a chance to earn additional reward points based on prediction accuracy:

- If the respondent's predicted value was within **±5%** of the true value, they would be eligible to receive **500 Japanese yen worth of reward points**.

Additional details shown to respondents:

- As an illustrative example (using an unrealistically large number), if the true value were **100,000 steps**, then any prediction between **95,000 and 105,000** would qualify.
- The ±5% bounds were calculated as integers by **rounding the upper bound up** and **rounding the lower bound down**.
- Eligible respondents would be notified by email by around **mid-March**.
- If many respondents qualified, a fair lottery would be conducted, and up to **24 respondents** (out of **1,200** survey respondents) would receive the additional reward.

**\*Additional Belief-Elicitation Tasks for Other Outcomes**

In addition to predicting average daily step counts, respondents completed analogous belief-elicitation tasks for subjective health status, prosocial attitudes, participation (take-up) rates, and favorability toward the implementing organization.

For each outcome, respondents were provided with benchmark information from the field experiment or the survey experiment and were asked to predict the corresponding outcome under the message + donation points condition. The same incentive rule (±5% accuracy threshold) was applied to all belief-elicitation tasks.